\documentclass[12pt]{article}
\usepackage{color}
\usepackage{amsmath, amssymb}
\usepackage{epsfig, palatino}
\usepackage{pstricks,pst-node,pst-tree}
\usepackage{epic}

\usepackage{mathrsfs}

\usepackage{ae} 
\usepackage[T1]{fontenc}
\usepackage[ansinew]{inputenc}
\usepackage{amsmath}
\usepackage{amssymb}
\usepackage{graphicx}
\usepackage{color}
\definecolor{darkblue}{cmyk}{0.9,0.9,0,0}
\usepackage[colorlinks=true,linkcolor=darkblue,citecolor=darkblue,urlcolor=darkblue]{hyperref}

\newcommand{\beq}{\begin{equation}}
\newcommand{\eeq}{\end{equation}}
\newcommand\beqa{\begin{eqnarray}}
\newcommand\eeqa{\end{eqnarray}}
\newcommand\bea{\begin{array}}
\newcommand\eea{\end{array}}

\def\XXint#1#2#3{{\setbox0=\hbox{$#1{#2#3}{\int}$}
\vcenter{\hbox{$#2#3$}}\kern-.5\wd0}}

\newcommand{\nn}{\nonumber}

\newcommand{\COMMENT}[1]{}

\newcommand{\neqa}{\nonumber\end{eqnarray}}
\newcommand{\la}[1]{\label{#1}}

\renewcommand{\d}{\partial}

\newcommand{\<}{{\langle}}
\renewcommand{\>}{{\rangle}}

\newcommand{\re}{\relax{\rm I\kern-.18em R}}

\def\su2{{SU(2)}}

\def\[{\left[}
\def\]{\right]}

\def\e{\epsilon}

\def\({\left(}
\def\){\right)}
\def\[{\left[}
\def\]{\right]}

\def\<{\langle}
\def\>{\rangle}

\def\pint{-\hskip-0.41cm \int}
\def\i2{\frac{i}{2}}

 \def\nref#1{{(\ref{#1})}}

\usepackage{varioref}
\usepackage{makeidx}
\makeindex
\usepackage[english]{babel}

\begin{document}

%

\thispagestyle{empty}

\renewcommand{\thefootnote}{\fnsymbol{footnote}}
\setcounter{footnote}{0}
\setcounter{figure}{0}
\begin{center}
$$$$
{\Large\textbf{\mathversion{bold}
Pulling the straps of polygons}\par}

\vspace{1.0cm}

\textrm{Davide Gaiotto${}^a$, Juan Maldacena$^a$, Amit Sever$^b$, Pedro Vieira$^b$}
\\ \vspace{1.2cm}
\footnotesize{

\textit{$^{a}$ School of Natural Sciences,\\Institute for Advanced Study, Princeton, NJ 08540, USA.} \\
\texttt{} \\
\vspace{3mm}
\textit{$^b$
Perimeter Institute for Theoretical Physics\\ Waterloo,
Ontario N2J 2W9, Canada} \\
\texttt{}
\vspace{3mm}
}


\par\vspace{1.5cm}

\textbf{Abstract}\vspace{2mm}
\end{center}

\noindent

Using the Operator Product Expansion for Wilson loops we derive a
simple formula giving the discontinuities of the two loop result in
terms of the one loop answer. We also argue that the knowledge of these
discontinuities should be enough to fix the full two loop answer, for a
general number of sides.
We work this out explicitly for the case of the hexagon and rederive
the known result.

\vspace*{\fill}

\setcounter{page}{1}
\renewcommand{\thefootnote}{\arabic{footnote}}
\setcounter{footnote}{0}

\newpage
\tableofcontents

\section{Introduction}

Scattering amplitudes in ${\cal N}=4$ super Yang Mills are related to
polygonal Wilson loops with null edges \cite{AmplitudeWilson}. The conformal symmetry on the Wilson loop
side corresponds to non-trivial integrability charges on the amplitude side.
For the case of ordinary operators, the conformal symmetry leads to the Operator Product Expansion.
For Wilson loops we have a similar expansion, whose systematics was explained in \cite{OPEpaper}.

In this paper we use this expansion to predict the form of certain discontinuities of the two loop
expression for Wilson loop correlators. Namely, the Wilson loop correlators are functions which
contain logarithmic branch cuts, corresponding to branch cuts at multi particle production thresholds for
scattering amplitudes. Some of the discontinuities of the amplitude are determined by the operator
product expansion. In the weak coupling perturbative expansion, this discontinuity can be obtained
from lower order results plus the knowledge of certain anomalous dimensions.

In the case of $\mathbb{R}^{1,1}$ kinematics we showed in \cite{Gaiotto:2010fk}
 that we could determine the two loop
results \cite{Heslop:2010kq}
 from the one loop results plus the anomalous dimensions.
In this paper we consider the general case of  full $\mathbb{R}^{1,3}$ kinematics and we derive an expression
for the discontinuities of the answer. We further conjecture that these are all the discontinuities
of the two loop answer. The expression we derive for the discontinuity has the form of an integral
kernel applied to the one loop result. We have not been able to do these integrals in general, however
we have shown that one can rederive the known result \cite{Anastasiou:2009kna,DelDuca:2009au,Volovich}
for the hexagon using this method.

The discontinuities are particularly simple at two loops. The reason is that the one loop result
comes from the tree level exchange of a gluon between different sides of a Wilson loop.
Thus, the OPE at one loop contains only single particle exchanges. These particles behave as free
particles and transform under simple representations of the full conformal group. At two loops
we get a logarithmic term in the OPE which comes from the anomalous dimension of these states.
These anomalous dimensions were computed in \cite{Basso}.
This logarithmic term gives rise to a discontinuity of the two loop answer. In fact, if we include
all states in the OPE expansion, we can multiply each term by the corresponding anomalous dimension and
perform the whole sum. This gives us an expression for the discontinuity. This whole operation reduces
to a simple kernel acting on the one loop answer. This is one of the main results of the paper.

In addition, we observe that these discontinuities appear to be enough to determine the answer.
This can be most simply explained by using the ``symbol'' operation introduced in \cite{Volovich}.
Namely the functions appearing in the two loop answer can be assigned a symbol which is a sum of
terms of the form
$ {\rm Sym}[A]= \sum R_1 \otimes R_2 \otimes R_3 \otimes R_4$. We conjecture that the first term
in the symbol is always a distance between two cusps. This is simply saying that we expect a
discontinuity whenever two cusps, which were originally spacelike separated, become null separated.
The number of independent discontinuities is the same as the number of independent OPE expansion
channels that we can have. Our expressions for the discontinuity  are transcendentality three functions which  determine the last three factors of
the symbol. Thus, if we had an efficient way to determine the symbol of our integral expression, we
could determine the symbol of the answer.

As we mentioned above, we have carried out this program for the simplest case of the hexagon.
In this way we have rederived the known answer for the hexagon. Of course, in doing so, we have
explicitly shown that the known two loop expression has a proper OPE expansion, in agreement with
the general discussion in \cite{OPEpaper}.

This paper is organized as follows.

In section two, we give
 a brief review of symbols and their properties since we found them useful
for stating our results and also for deriving some of them. We also explain why the
various OPE channels determine the symbol of the two loop answer, if one assumes that the
first entry in the symbol is a distance.
In section three we compute the OPE expansion at one loop. Namely, we write the part of
the one loop answer that contributes to the OPE. This is given by a correlation  function of two
Wilson loops. We then decompose this answer into the contribution from self dual and anti-self dual
fields. This is necessary since the anomalous dimensions will treat these two cases differently.

In section four we describe the various states that are propagating in the one loop result.
We  write their anomalous dimensions, originally derived in \cite{Basso}.
We also show that we can describe the action of the anomalous dimension as
certain convolution kernel.

In section five we describe the hexagon.
We perform the expansions explicitly and we compute the action of the
convolution kernel indirectly, by considering the symbol and
the possible discontinuities. This is a method
that could be useful in the general case, though here we have only
derived it for the case of the hexagon.

Some technical details are relegated to an appendix. In addition, we have two appendices
with two special kinematic limits where the answer simplifies and can be computed more easily.

\section{The OPE and the Symbols}\la{rev}

We have found that the notion of a ``symbol'' introduced in \cite{Volovich} (and references therein) is
very useful for dealing with the type of transcendental functions that appear in these Wilson loop computations.
Our final answer can be stated without any reference to the symbol. However, it was useful to use
symbols for intermediate steps, as well as for thinking about these functions. Thus we will begin with a
quick review of symbols.

\subsection{Introduction to symbols}\la{sym}

Transcendental functions such as polylogarithms and their generalizations
satisfy many intricate identities. This implies that given two combinations of polylogarithms,
 it can be very tricky to recognize if they coincide or not.
There is a linear operation on transcendental functions, which we will denote as a map
$F \to \text{Sym}[F]$, which is very useful to keep track of all sort of polylogarithm identities.
The polylogarithm identities satisfied by a function $F$ become trivial algebraic identities
satisfied by its ``symbol'' $\text{Sym}[F]$. The symbol $\text{Sym}[F]$ loses some of the information about $F$,
but it is still very useful.

Functions of transcendentality degree $n$ are defined recursively, as functions
with logarithmic cuts, such that the discontinuity across the cuts is $2 \pi i$ times
functions of transcendentality degree $n-1$. Constants (or rational functions), by definition, have degree $0$.
We can give some simple examples. The logarithm $\log x$ has degree $1$: is has a cut running along the negative real axis, with discontinuity $2 \pi i$.
The dilogarithm $Li_2(x)$ has degree $2$: It has a cut running from $1$ to $\infty$ with discontinuity $- 2 \pi i \log x$.

The symbol of a function of transcendentality $n$ is a linear combination of elements of the form
\begin{equation}
R_1 \otimes R_2 \otimes R_3 \otimes R_4 \cdots \otimes R_n
\end{equation}
 where the $R_i$ are rational functions. It should be thought of as a tensor product of logarithms
 \begin{equation}
\log R_1 \otimes \log R_2 \otimes \log R_3 \otimes \log R_4 \cdots \otimes \log R_n
\end{equation}
but the ``$\log$'' is usually omitted.
 We have the following recursive definition:
 \begin{equation}
 \text{Sym}[F] = \sum_a R_a \otimes \text{Sym}[F_a]
 \end{equation}
if $F$ has logarithmic cuts which start at $R_a=0$, end at $R_a=\infty$, with discontinuity $2 \pi i F_a$.

For example, $\text{Sym}[1]=0$, $\text{Sym}[\log x] = x$, $\text{Sym}[-Li_2(x)] = (x-1) \otimes x$. The symbol of $\log x \log y$ is $x \otimes y + y \otimes x$: it has discontinuity $2 \pi i \log y$ around $x=0$ and
$2 \pi i \log x$ around $y=0$. Notice that the difference of two functions of degree $n$ with the same symbol
can only be a function which becomes $0$ after taking $n$ discontinuities, i.e. a function of transcendentality degree $n-1$ or lower.
This is the amount of information which is lost by taking the symbol.

From the definition, a lot of useful properties follow right away. The most important are that  the symbol behaves like a tensor product of logarithms
\begin{equation}
\cdots \otimes x\,y \otimes \cdots =  \cdots \otimes x \otimes \cdots  +  \cdots \otimes y \otimes \cdots
\end{equation}
and it is transparent to constants:
 \begin{equation}\cdots \otimes (c R_1) \otimes \cdots = \cdots \otimes R_1 \otimes \cdots \end{equation}
if $c$ is a constant.
We can illustrate this for a simple dilogarithm identity:
\begin{equation}
Li_2(x) + Li_2(1/x) = -\frac{\pi^2}{6} - \frac{1}{2} \log(-x)^2\nn
\end{equation}
becomes, upon taking the symbol,
\begin{equation}
- (1-x) \otimes x - (1-1/x) \otimes (1/x) = - (1-x) \otimes x + {x-1\over x} \otimes x = (-1/x) \otimes x= - x \otimes x\nn
\end{equation}

The symbol of the product of two functions $\text{Sym}[F G]$ can be built from the relation on discontinuities $\Delta (FG) = (\Delta F) G + F \Delta G$.
By linearity of the tensor product, we only need to
compute it for $\text{Sym}[F]=\otimes_{i=1}^n R_i$ and $\text{Sym}[G]=\otimes_{i=n+1}^m R_i$:
 \begin{equation}
 \text{Sym}[F G] = \sum_\pi \otimes_{i=1}^{n+m} R_{\pi(i)}
 \end{equation}
 where the sum is over all the permutations of the full set of $n+m$ $R_i$ which preserve the original ordering
  among the factors in $\text{Sym}[F]$ and among the factors in $\text{Sym}[G]$.

  For example, if $n=m=2$,
 \begin{align}
 \text{Sym}[F G] =& R_1 \otimes R_2 \otimes R_3 \otimes R_4 + R_1 \otimes R_3 \otimes R_2 \otimes R_4 +  R_1 \otimes R_3 \otimes R_4 \otimes R_2 + \\&+  R_3 \otimes R_1 \otimes R_2 \otimes R_4 + R_3 \otimes R_1 \otimes R_4 \otimes R_2 +  R_3 \otimes R_4 \otimes R_1 \otimes R_2      \end{align}
Hence we can easily build the symbol of any polynomial in polylogarithms and logarithms.

There is a relation between the symbol of a function and the definition through iterated integrals.
Indeed, the operation of taking a discontinuity commutes with differentiation,
hence the symbol of a derivative $\text{Sym}[dF]$ is given by the replacement in $\text{Sym}[F]$
\begin{equation}
R_1 \otimes \cdots \otimes R_n \to d\log R_n \left[ R_1 \otimes  \cdots \otimes R_{n-1}\right]
\end{equation}
Not all possible symbols are symbols of a function. For example, if $x$ and $y$ are unrelated,
$x \otimes y$ is not the symbol of a function. A useful test to check if a symbol could be the symbol of a function is to
pick two consecutive slots of the symbol, and replace  \cite{Goncharov}
\begin{equation} \label{testsymb}
R_1 \otimes \cdots \otimes R_i \otimes R_{i+1} \otimes  \cdots \otimes R_n \to d\log R_i \wedge d\log R_{i+1} \left[ R_1 \otimes \cdots \otimes  \cdots \otimes R_n \right]
\end{equation}
This transformation maps the symbol of functions to zero.

\subsection{On OPE constructibility}
\label{OPEconstr}

The null polygonal Wilson loops which we study are defined as the limit of Wilson loops with space-like
sides, and space-like separation between any pair of points. In the limit, we make the sides of the polygon null, but
focus on the kinematic region where all points which are not on the same side remain space-like separated.
Infrared divergencies which arise in the null limit are well understood. The answer should be smooth in this kinematic region.
The kinematic region of interest is bounded by walls where
two non-consecutive vertices become null-separated, i.e. $(x_i -x_j)^2=0$. We can analytically continue the amplitude around these loci, but
the analytically continued answer will generically have cuts which end at these loci.

At one loop, the Wilson loop amplitude is a function of transcendentality degree two. It has logarithmic cuts which start at $(x_i -x_j)^2=0$.
More precisely, if we take the symbol of the one-loop answer, it basically takes the form
\begin{equation}
S^{\mathrm{1-loop}}=\sum_{ij} (x_i -x_j)^2 \otimes {\rm Sym}[D^{\mathrm{1-loop}}_{ij}]
\end{equation}
where $D^{\mathrm{1-loop}}_{ij}$ are the   discontinuities at the corresponding cuts.

The two-loop answer will in general also have cuts which start at the loci $(x_i -x_j)^2=0$. In order to bootstrap the two-loop amplitude,
we will make a crucial assumption:  we will require the two loop answer to be a function of transcendentality degree four, which has
transcendentality degree three discontinuities only across the cuts starting at the the loci $(x_i -x_j)^2=0$. More precisely, we will require the symbol of the two-loop answer to take the form
\begin{equation} \label{twoloopsymb}
S^{\mathrm{2-loop}}=\sum_{ij} (x_i -x_j)^2 \otimes {\rm Sym}[D^{\mathrm{2-loop}}_{ij}]
\end{equation}
We will now show how to compute the discontinuities around the loci $(x_i -x_j)^2=0$ by an analysis of OPEs and hence compute the full symbol of the two loop amplitude.
Notice that a generic formal symbol is not the symbol of an actual function. Given the set of ${\rm Sym}[D^{\mathrm{2-loop}}_{ij}]$, we can check if $S^{\mathrm{2-loop}}$ in \nref{twoloopsymb}
is the symbol of a function or not.
If it is, our basic assumption passes a very strong self-consistency check. If it is not, it means that the two loops amplitude does not have a symbol or alternatively, that our original set of discontinuities was incomplete, and
the amplitude has other cuts further away from the original kinematic region, which need to be accounted for in the boot-strap procedure.

\subsubsection{Discontinuities from OPEs}

The main idea behind OPEs is to study (multi)collinear limits of the amplitude, by acting on a portion of the Wilson loop with a one-parameter family of conformal transformations
$M_{\tau}$, chosen in such a way that as $\tau \to \infty$ that portion of the Wilson loop is pushed towards a (multi) collinear limit \cite{OPEpaper}.
The distances between the vertices in that portion of the Wilson loop scale uniformly as $e^{- \tau}$ for large $\tau$. The family of Wilson loops remains
 inside the preferred Euclidean kinematic region, and approaches the boundary as $\tau \to \infty$. An OPE expansion channel is chosen by selecting a pair of null sides separated by
two or more other null sides. We call one of the sides the ``left'' side and the other the ``right'' side.

The general analysis in \cite{OPEpaper}
 constrains the behavior of the amplitude at large $\tau$ by an OPE analysis, which replaces the portion of Wilson loop by
a sum of operator insertions on the $\tau \to \infty$ limit of the Wilson loop.
At two loops, the result is an expansion for the amplitude of the form
\begin{equation}
b_0 + (a_1 \tau + b_1) e^{- \tau} + (a_2 \tau + b_2) e^{- 2 \tau}+(a_3 \tau + b_3) e^{-3 \tau}+\cdots
\end{equation}
If we complexify $\tau$, we can choose a path in the space of Wilson loops in order to compute useful discontinuities: go to very large $\tau$,
then take the discontinuity of the amplitude under the shift $\tau \to \tau + 2 \pi i$, which is a small closed loop in the space of Wilson loops.
 Upon taking this discontinuity, only the $\tau e^{- n \tau}$ terms contribute. These are very special:
they arise from the ``anomalous dimension'' of the operators, i.e. the leading correction to their quantum numbers under the $M_\tau$ transformation.
Hence the OPE discontinuity of the two-loop amplitude differs from the full one-loop amplitude only by the insertion of this anomalous dimension operator.

This path at large $\tau$,
 winds once around all the loci $(x_i -x_j)^2=0$ for all pairs of vertices $i,j$ in the portion of the path acted upon by $M_\tau$.
This ``OPE discontinuity''  is
\begin{equation}
D_{L,R} = \sum_{ij}D^{\mathrm{2-loop}}_{ij}
\end{equation}
where the sum runs over all pairs of vertices $i,j$ in the portion of the path acted upon by $M_\tau$. These are
the vertices which are between the left and right sides as defined above.

This means that we can reconstruct all the $D^{\mathrm{2-loop}}_{ij}$ recursively. At the first step, we extract
$D^{\mathrm{2-loop}}_{i, i+2}$ from the OPE discontinuity corresponding to the collinear limit of the $i,i+1,i+2$ vertices.
At the second step, we extract $D^{\mathrm{2-loop}}_{i, i+3}$ from the OPE discontinuity corresponding to the collinear limit of the $i,i+1,i+2, i+3$ vertices, which is
\begin{equation}
D_{L,R} = D^{\mathrm{2-loop}}_{i, i+3} + D^{\mathrm{2-loop}}_{i, i+2}+ D^{\mathrm{2-loop}}_{i+1, i+3}
\end{equation}
and so on.

Thus, if we have a way to computing all the OPE discontinuities for all OPE channels, namely, for any
choice of two sides $L,R$, we can then construct the symbol for the two loop answer
\nref{twoloopsymb}. Below we will derive an expression for this discontinuity, as a function.
In fact,
 $D_{L,R}$, and thus, also $D_{ij}^{\mathrm{2-loop}}$ are symbols of a function. This also follows
  from the fact that we can vary the $X_i . X_j$ independently.
  Note that
 conformal invariance
implies  relations among the $D_{ij}^{\mathrm{2-loop}} =  D_{ji}^{\mathrm{2-loop}}$, and
 $\sum_{i} D_{ij}^{\mathrm{2-loop}} =0$ for each $j$.

\section{The one loop sources} \la{sec3}

In order to compute the two loop discontinuities we need to compute the one loop contribution to
the OPE. This is simply the one loop result for the Wilson loop where we keep only the part that depends
non-trivially on the OPE  ``time'' parameter $\tau$. This is completely captured by a single particle
exchange between the ``top '' part of the polygon and the ``bottom'' part. The top and bottom parts are
separated by the two special sides, called ``left'' and ``right''. Selecting and OPE channel is the same
as selecting these two special sides. This can be derived from \cite{Bern:2005iz}. We rederive these
results from scratch here.

\subsection{Short review on kinematics}

We represent $\mathbb{R}^{1,3}$ as the lightcone in $\mathbb{R}^{2,4}$. That is, any point $x\in \mathbb{R}^{1,3}$ is associated with a null ray in $\mathbb{R}^{2,4}$, $\{X\in \mathbb{R}^{2,4},\ X^2=0,\ X\simeq tX\}$.
The map to the usual Poincare coordinates is
\beq
x_\mu={X_\mu\over X_{-1}+X_4}\ .
\eeq
The conformal group is then realized linearly as the $SO(2,4)$ rotations of the embedding coordinates (the $X$'s). A convenient way of representing a null polygon is using momentum twistors \cite{Hodges:2009hk}.
These are $\mathbb{R}^{2,4}$ spinors that are defined up to rescaling $\lambda=(\lambda_1,\lambda_2,\lambda_3,\lambda_4)\simeq t\lambda$. The conformal group acts on these by a simple $SL(4)$ left group multiplication. With any pair of momentum twistors $\lambda,\tilde\lambda$ we can associate a null ray in $\mathbb{R}^{2,4}$ and therefore a point $x\in \mathbb{R}^{1,3}$ as
\beq \label{PointDef}
X_{ab}=X_M\Gamma^M_{ab}=\lambda_{[a}\tilde\lambda_{b]}\ ,\qquad \begin{array}{l}M=-1,0,1,2,3,4\\ a,b=1,2,3,4\end{array}\ ,
\eeq
where $\Gamma_{M}$ are sigma matrices in $\mathbb{R}^{2,4}$. A polygon with $N$ null sides can be given as a sequence of $N$ twistors $\lambda_i$, such that the intersection
 of the sides $i$ and $i+1$ is the point $X_{i } = \lambda_i \wedge \lambda_{i+1}$. The distance between two cusps of the polygon reads
\beq\la{distance}
(x_i-x_j)^2={\lambda_i\wedge\lambda_{i+1}\wedge\lambda_j\wedge\lambda_{j+1}\over
(\lambda_i^T\cdot\Gamma^+\cdot\lambda_{i+1})(\lambda_j^T\cdot\Gamma^+\cdot\lambda_{j+1})}\ ,
\eeq
where $\Gamma^+=\frac{i}{\sqrt{2}}\(\Gamma^{-1}+\Gamma^4\)$.
The numerator in \nref{distance} denotes the SL(4) invariant product of four twistors, given by contracting
 with the $\epsilon^{a_1 \cdots a_4}$ tensor of $SL(4)$.
 We automatically have $X_{i-1 } \cdot X_{i }=0$,
which
is the condition that the $i$-th side should be null. It is also useful to introduce dual momentum twistors $\hat \lambda_i$, which can be defined as $\hat \lambda_i =
 \lambda_{i-1} \wedge \lambda_i \wedge \lambda_{i+1}$. They also satisfy  $X_{i } = \hat \lambda_i
 \wedge \hat \lambda_{i+1}$ and transform by a right $SL(4)$ multiplication.  A twistor $\lambda$ together with an orthogonal dual twistor $\hat \mu $, $(\hat \mu, \lambda)= \lambda_a \hat \mu^a = 0$,
 defines a null line (and viceversa). The line is given by
 all points of the form $Z = \lambda \wedge v$
 with $(\hat \mu,v)=0$.

\subsection{One-loop pairing}

\begin{figure}[t]
\centering
\def\svgwidth{13cm}
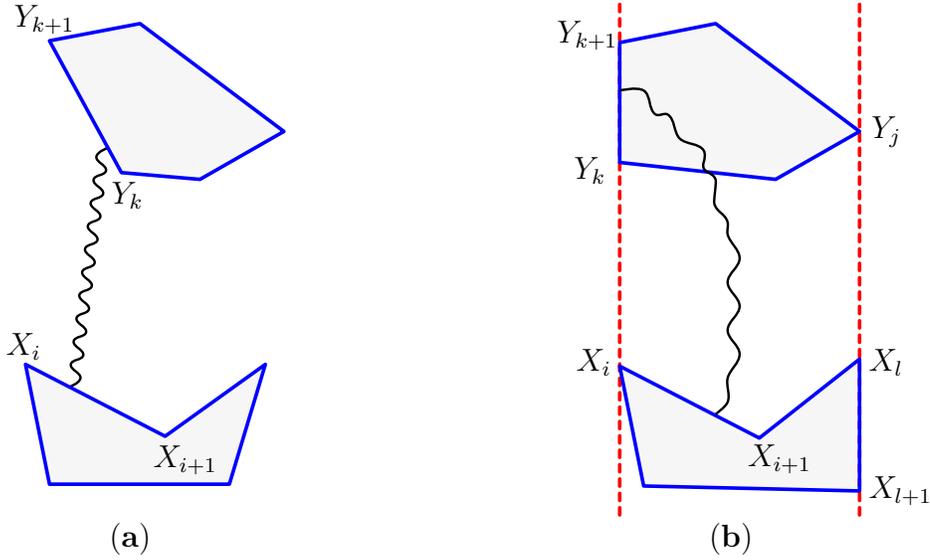
\caption{  Correlation function between two Polygonal Wilson loops, which is given by a single gluon
exchange. In (a) we consider a Polygon where all the points in the $X$ contour are spacelike
separated from all points in the $Y$ contour. (b) We take a limit of (a) where
a segment of one contour $(Y_k,Y_{k+1})$ is on the same null line as a cusp of the other contour $X_i$. The rest
 of the points are spacelike separated.}\label{Wilsoncorrelator}
\end{figure}

In this section we compute the integral of the photon propagator
stretched between a pair of closed null-polygonal Wilson loops with relative space-like separations (see figure \ref{Wilsoncorrelator}).

The calculation is in principle straightforward. We will use a six-dimensional description of conformal space-time.
Points in spacetime correspond to equivalence classes of null six-dimensional vectors $X$ up to rescaling. A possible
choice of representative takes the form $X=(1,x^\mu, x^2)$, so that $X \cdot Y = (x-y)^2$.
We will pick two six dimensional representatives $X(s)$, $Y(s')$ for the paths.
Then the integral
\begin{equation}
r_{U(1)}(X,Y)=\oint \oint \frac{dX \cdot dY}{X\cdot Y}
\end{equation}
will do the job. The rescaling invariance of the integral is a bit subtle: a rescaling $X(s) \to \lambda(s) X(s)$ will shift the integrand by
$d_s\lambda d_{s'}\log X\cdot Y$. As long as the polygons are space-like separated, so that the $\log X\cdot Y$ has no monodromy,
the integral is conformal invariant. In the conformal gauge where $X\cdot Y$ coincides with the four-dimensional squared distance,
the integrand $\frac{dX \cdot dY}{X\cdot Y}$ reduces to the desired $\frac{dx \cdot dy}{(x-y)^2}$.

The integral can be done by brute force, say by picking the parameterization $X(s) = (1-s) X_i + s X_{i+1}$ in the $i$-th side of the first polygon, and similarly for the second polygon.
The result of the side-by-side brute force calculation is a sum of dilogarithms and bilinears of logarithms, but the arguments of the dilogarithms
contain linear combinations of the $X_i \cdot Y_j$ which are not well-behaved under rescaling. Dilogarithm identities need to be used in order to eliminate this spurious dependence.
A shortcut to deal with such identities is to compute the symbol of the answer. We refer to section \ref{sym} for a review of symbols of polylogarithms.

The integral $r_{U(1)}(X,Y)$ has interesting discontinuities around the loci where the polygons become null separated first.
It is not difficult to see that generically  this happens first when a vertex of a polygon becomes null separated from a vertex in the other polygon.
Indeed if a smooth point $X_0= X(s_0)$ for $X(s)$ is null separated from a point $Y_0$ of $Y(s')$, $X(s).(Y_0+ \delta Y) \sim (s-s_0) \dot X(s_0) \cdot Y_0 + X_0 \cdot \delta Y$
so either a whole side of $X$ is becoming null separated to $Y_0$ or we can find a solution for $s-s_0$ describing points of $X$ near $X_0$ null separated from points of $Y$ near $Y_0$.
The symbol of the full answer turns out to have the form
\begin{equation}
S =\text{Sym}[r_{U(1)}(X,Y)]= \sum_{i,k} X_i \cdot Y_k \otimes \text{Sym}[\Delta_{ik}].
\end{equation}
The integral has a logarithmic cut originating from each of the $X_i \cdot Y_k=0$ loci,
with a discontinuity $2 \pi i \Delta_{ik}$.

If two vertices $X_i$, $Y_k$ are close to be null separated, one can study directly the local behavior of the integral near the two
vertices. The result is that
\begin{align}
\Delta_{ik}= &\log \left[1- \frac{(X_i \cdot Y_k)( X_{i-1}\cdot Y_{k+1})}{(X_i \cdot Y_{k+1})( X_{i-1}\cdot Y_{k})}\right] +\log \left[1- \frac{(X_i \cdot Y_k)( X_{i+1}\cdot Y_{k-1})}{(X_i \cdot Y_{k-1})( X_{i+1}\cdot Y_{k})}\right] \notag \\ -& \log \left[1- \frac{(X_i \cdot Y_k )(X_{i-1}\cdot Y_{k-1})}{(X_i \cdot Y_{k-1})( X_{i-1}\cdot Y_{k})}\right]-\log \left[1- \frac{(X_i \cdot Y_k)( X_{i+1}\cdot Y_{k+1})}{(X_i \cdot Y_{k+1})( X_{i+1}\cdot Y_{k})}   \right]
\end{align}

We can now write down a function with symbol $S$, and no discontinuities in the region where the polygons have space-like separation. This should agree
with $r_{U(1)}(X,Y)$ up to a constant, which can be fixed simply by
requiring the answer to go to zero as the two polygons are brought far from each other.

We can rewrite $S$ as
\begin{align}
S=\sum_{ik} & \frac{(X_i \cdot Y_k)(X_{i-1} \cdot Y_{k+1})}{(X_i \cdot Y_{k+1})(X_{i-1} \cdot Y_k)} \otimes \left[1- \frac{(X_i \cdot Y_k)( X_{i-1}\cdot Y_{k+1})}{(X_i \cdot Y_{k+1})( X_{i-1}\cdot Y_{k})}\right] \notag \\ -& (X_i \cdot Y_k) \otimes \frac{(X_i \cdot Y_k )(X_{i-1}\cdot Y_{k-1})}{(X_i \cdot Y_{k-1})( X_{i-1}\cdot Y_{k})}\frac{(X_i \cdot Y_k)( X_{i+1}\cdot Y_{k+1})}{(X_i \cdot Y_{k+1})( X_{i+1}\cdot Y_{k})}
\end{align}
The first line
is the symbol of a sum of dilogarithms
\begin{equation} \label{roneu}
r_1=-\sum_{ik} Li_2\left[1- \frac{(X_i \cdot Y_k)( X_{i-1}\cdot Y_{k+1})}{(X_i \cdot Y_{k+1})( X_{i-1}\cdot Y_{k})}\right]
\end{equation}
and the second part can be reorganized a bit
to the symbol of the rescaling- invariant combination
 \begin{equation} \label{rtwou}
r_2=-\sum_{ik}  \log X_i \cdot Y_k \log \frac{(X_i \cdot Y_k)( X_{i+1}\cdot Y_{k+1})}{(X_i \cdot Y_{k+1})( X_{i+1}\cdot Y_{k}) }
\end{equation}
Indeed a rescaling of $X_i$ or of $Y_k$ shifts $r_2$ by a sum which telescopes to zero.
Then $r_{U(1)}(X,Y) = r_1 + r_2$ in \nref{roneu} and \nref{rtwou}.
Notice that in the kinematical region of space-like separation, we can take $X_i \cdot Y_k >0$, and the functions above have manifestly no cuts.
Notice also that the discontinuity function $\Delta_{ik}$ are zero at the boundary $X_i \cdot Y_k=0$, which means that the
$r_{U(1)}(X,Y)$ does not diverge if a vertex of $X$ and a vertex of $Y$ are brought to be null-separated.

There is a useful rearrangement of the answer in terms of momentum twistors. Momentum twistors are four-components vectors which transform in the fundamental of
the $SL(4)$ conformal group. A six-dimensional null-vector $X$ can always be written as the wedge product of two momentum twistors \nref{PointDef}.
For a generic pair of polygons, we can pick momentum twistors $\lambda_i$ associated to sides of $X$ and $\mu_k$ associated to sides of $y$, so that
$X_i =\lambda_i \wedge \lambda_{i+1}$ and $Y_k = \mu_k \wedge \mu_{k+1}$.
Then inner products are written as determinants, $X_i \cdot Y_k = (\lambda_i \wedge \lambda_{i+1} \wedge \mu_k \wedge \mu_{k+1})$ and we have a very useful Plucker identity:
starting for example with
\beqa
(X_i \cdot Y_{k+1})( X_{i-1}\cdot Y_{k})\!\!\!&-&\!\!\!(X_i \cdot Y_k)( X_{i-1}\cdot Y_{k+1})=  \\
 && (\lambda_i \wedge \lambda_{i+1} \wedge \mu_{k+1} \wedge \mu_{k+2}) (\lambda_{i-1} \wedge \lambda_{i} \wedge \mu_k \wedge \mu_{k+1})\nn\\&&- (\lambda_i \wedge \lambda_{i+1} \wedge \mu_k \wedge \mu_{k+1}) (\lambda_{i-1} \wedge \lambda_{i} \wedge \mu_{k+1} \wedge \mu_{k+2})\nn
\eeqa
we observe that $\lambda_i \wedge \lambda_{k+1}$ is common to all determinants. Modulo $\lambda_i$, $\lambda_{k+1}$ we can use a standard two-dimensional cyclic identity
and rewrite
\beqa
(X_i \cdot Y_{k+1})( X_{i-1}\cdot Y_{k})-(X_i \cdot Y_k)( X_{i-1}\cdot Y_{k+1})  &=& (\lambda_i \wedge \mu_k \wedge \mu_{k+1} \wedge \mu_{k+2}) (\mu_{k+1} \wedge \lambda_{i-1} \wedge \lambda_i \wedge \lambda_{i+1})\nn
\\
&=& ( \lambda_ i , \hat \mu_{k+1} ) ( \mu_{k+1} , \hat \lambda_i )
\eeqa
where we have introduced the dual momentum twistors for the contours, as
$\hat \lambda_i =  \lambda_{i-1} \wedge \lambda_i \wedge \lambda_{i+1}$ and $\hat \mu_{k+1} = \mu_k \wedge \mu_{k+1} \wedge \mu_{k+2}$.
 We can combine the four logs in $\Delta_{ik}$, the denominators drop out and we get simply the product of two twistor cross-ratios
\begin{equation}
\Delta_{ik} = \log \left[ \frac{(\lambda_i, \hat \mu_{k+1})(\lambda_{i+1}, \hat \mu_{k})}{(\lambda_i, \hat \mu_{k})(\lambda_{i+1}, \hat \mu_{k+1})}\frac{(\mu_{k+1},\hat \lambda_i)(\mu_{k},\hat \lambda_{i+1})}{(\mu_{k},\hat \lambda_i)(\mu_{k+1},\hat \lambda_{i+1})}\right]
\end{equation}

This is a very neat answer. Notice that the symbol $S$ of $r_{U(1)}(X,Y)$ can be now split into two pieces:
\begin{equation}
S^+ = \sum_{i,k} X_i \cdot Y_k \otimes \frac{(\lambda_i, \hat \mu_{k+1})(\lambda_{i+1}, \hat \mu_{k})}{(\lambda_i, \hat \mu_{k})(\lambda_{i+1}, \hat \mu_{k+1})} = X_i \cdot Y_k \otimes \Delta^+_{ik}
\end{equation}
and
\begin{equation}
S^- = \sum_{i,k} X_i \cdot Y_k \otimes \frac{(\mu_{k+1},\hat \lambda_i)(\mu_{k},\hat \lambda_{i+1})}{(\mu_{k},\hat \lambda_i)(\mu_{k+1},\hat \lambda_{i+1})} = X_i \cdot Y_k \otimes \Delta^-_{ik}
\end{equation}

We can even write down separate functions $r^\pm$ with these two symbol. The crucial observation is that
\begin{equation}
 X_i \cdot Y_k = (\lambda_i, \hat \mu_{k})(\lambda_{i+1}, \hat \mu_{k+1})-(\lambda_i, \hat \mu_{k+1})(\lambda_{i+1}, \hat \mu_{k})=(\mu_{k},\hat \lambda_i)(\mu_{k+1},\hat \lambda_{i+1}) - (\mu_{k+1},\hat \lambda_i)(\mu_{k},\hat \lambda_{i+1})\nn
\end{equation}
Hence we can subtract off from $S^+$
\begin{equation}
S^+_1 = \sum_{i,k} \left[ 1- \frac{(\lambda_i, \hat \mu_{k+1})(\lambda_{i+1}, \hat \mu_{k})}{(\lambda_i, \hat \mu_{k})(\lambda_{i+1}, \hat \mu_{k+1})} \right]\otimes \frac{(\lambda_i, \hat \mu_{k+1})(\lambda_{i+1}, \hat \mu_{k})}{(\lambda_i, \hat \mu_{k})(\lambda_{i+1}, \hat \mu_{k+1})}
\end{equation}
which is the symbol of
\begin{equation} \label{roneplus}
r^+_1 = - \sum_{i,k} Li_2\left[ \frac{(\lambda_i, \hat \mu_{k+1})(\lambda_{i+1}, \hat \mu_{k})}{(\lambda_i, \hat \mu_{k})(\lambda_{i+1}, \hat \mu_{k+1})}\right]
\end{equation}
and be left with the symbol of a bunch of logarithms
\begin{equation} \label{rtwoplus}
r^+_2 =  \sum_{i,k} \log (\lambda_i, \hat \mu_{k}) \log \frac{(\lambda_i, \hat \mu_{k+1})(\lambda_{i+1}, \hat \mu_{k})}{(\lambda_i, \hat \mu_{k})(\lambda_{i+1}, \hat \mu_{k+1})}
\end{equation}
Thus $r^+ = r_1^+ + r_2^+$ in \nref{roneplus}, \nref{rtwoplus}. $r^-$ is given by the same expressions
exchanging twistors and co-twistors, namely $\lambda_i \to \hat \lambda_i$, $\hat \mu_k \to \mu_k$. And
finally, $r_{U(1)} = r^+ + r^- $.
Actually, if we want to have answers which explicitly have no cuts in the region of space-like separation of the two
loops we should pick the arguments of the dilogarithms with a little more care. The boundaries $X_i \cdot Y_k=0$
are the locus where the argument of the corresponding dilogarithm becomes 1.
In a region of space-like separation, some arguments will be strictly larger than $1$ and some strictly smaller than $1$. Indeed
the product of the arguments over $k$ at fixed $i$ telescopes to $1$. In any component of the region of space-like separation
we should do some replacements $Li_2(1/x) \to -Li_2(x) - \frac{1}{2} \log^2 x$ so that all dilogarithms have no cuts on the correct side of $1$.

For our computation of the  two loops remainder function we have to consider the case where a cusp of the bottom polygon $X_i$ is null separated from an edge of the top polygon $(Y_k,Y_{k+1})$ and a cusp of the top polygon $Y_j$ is null separated from an edge of the bottom polygon $(X_l,X_{l+1})$,
see figure \ref{Wilsoncorrelator}(b). In this case, $r^\pm$ reduce to   manifestly finite functions.
This can be found explicitly in appendix \ref{limitingvalues}.

In order to understand better the meaning of $S^\pm$ and the corresponding functions $r^\pm(X,Y)$, it is useful to review a bit of twistorial geometry.

\subsection{Null planes and null lines}
In $(2,2)$ signature it is possible to find closed null triangles. Given three momentum twistors $\lambda_1$, $\lambda_2$, $\lambda_3$ we can make a triangle the standard way,
with vertices $X_i = \lambda_i \wedge \lambda_{i+1}$. The whole triangle satisfies the constraint $\hat \lambda \cdot X=0$, with the dual momentum twistor $\hat \lambda$
defined as $\hat \lambda = \lambda_1 \wedge \lambda_2 \wedge \lambda_3$. This is a rather interesting equation. It defines a null plane, which we will denote as $\hat S$.
Two such null planes either coincide, or intersect at a single point: the equations $\hat \lambda \cdot X=0$ and $\hat \lambda' \cdot X=0$ have a unique solution (up to rescaling) for $X$.

There is a second type of null planes, where we exchange the roles of twistors and dual twistors. They are planes containing the second type of possible null triangles:
triangles with vertices of the form $X_i = \lambda \wedge \lambda_i$ for some $\lambda$. The points $X = \lambda \wedge v$ for some fixed $\lambda$, variable $v$
generate a null plane which we will indicate as $S$. We can also describe it as the space of solutions to the equation $X \wedge \lambda=0$.
Again, two null planes $S$ and $S'$ either coincide or meet at a unique point, $X=\lambda \wedge \lambda'$.

A $S$ plane and a $\hat S$ plane with $\lambda \cdot \hat \lambda \neq 0$ will not intersect at all: $\hat \lambda \cdot (\lambda \wedge v)$ is never zero.
On the other hand, if $\lambda \cdot \hat \lambda = 0$ the two planes meet along a null line: the set of points such that $X = \lambda \wedge v$ with $\hat \lambda \cdot v=0$.
Vice-versa, any null line can be described this way, in terms of a twistor and a dual twistor:
given two null-separated points $X_1$ and $X_2$ we can always write $X_1 = \lambda \wedge \lambda_1$ and $X_2 = \lambda \wedge \lambda_2$
and then define a $\hat \lambda = \lambda \wedge \lambda_1 \wedge \lambda_2$. The two points and any linear combinations of them lie on the line defined by $\lambda, \hat \lambda$.
So a pair of null-separated points define two null planes of opposite type and a null line through them.

Given a null plane $\hat S$ and a point $X$ not on it, the twistor $\lambda = \hat \lambda \cdot X$ defines a second null plane $S$ which passes through $X$ and intersects $\hat S$ along a null line.
The same is true for a null plane $S$ and a point $X$ not on it, though then it is $\hat \lambda = X \wedge \lambda$ that defines a second null plane $\hat S$ and null line.
Given a line $\ell$ parameterized by a pair $(\lambda, \hat \lambda)$ and a point $X$ away from the line, we can find a unique point on the line which is null separated from $X$:
$X_\ell = (\hat \lambda \cdot X) \wedge \lambda$. The null line through $X$ and $X_\ell$ is given by the pair $(\hat \lambda \cdot X, X\wedge \lambda)$.

\subsection{Helicity interpretation of $r^\pm(X,Y)$}
It is interesting to compute the pairing $r_{U(1)}(X,Y)$ of two null triangles of the same type.
If we look at two triangles of the same type, say with vertices $X_i = \lambda \wedge \lambda_i$ and $Y_k = \tilde \lambda \wedge \tilde \lambda_k$,
we notice that any inner product $X_i \cdot Y_k = \lambda \wedge \lambda_i \wedge \tilde \lambda \wedge \tilde \lambda_k$ involves the
quantity $\lambda \wedge \tilde \lambda$, hence depends on $\lambda_i$,$\tilde \lambda_k$ only through their value modulo $\lambda, \tilde \lambda$.
This reduces the cross-ratios effectively to 2d cross-ratios, and it is easy to check that $\Delta_{ik}=0$. Hence $r_{U(1)}(X,Y)$ (or at least its symbol) vanishes for two triangles of the same type.
This has a simple physical interpretation: the integral of the $U(1)$ gauge field along a triangular contour is the same as the integral of the field strength
on a surface bounded by the triangle. If the triangle is null, it lies on a null plane and we can take the surface to lie on the plane as well.
But the surface elements of the two types of null-planes are self-dual and anti-self-dual respectively, and hence couple to photons of positive or negative helicity
only. But the propagator is non-zero only for zero total helicity.

The pairing $r_{U(1)}(X,Y)$ for triangles of opposite type is a non-trivial function of the three independent cross ratios which can be built from the nine inner products
$\lambda_i \cdot \hat \mu_k$ of the three twistors which define a triangle, and the three dual twistors which define the other (or vice-versa, depending on the type of triangles).
The symbol $S$ can be computed as done for general polygons, but it differs on one important way: it only includes
one of the two twistorial cross-ratios, hence $S=S^+$ (or $S=S^-$, depending on the type of triangles).

\begin{figure}[t]
\centering
\def\svgwidth{15cm}
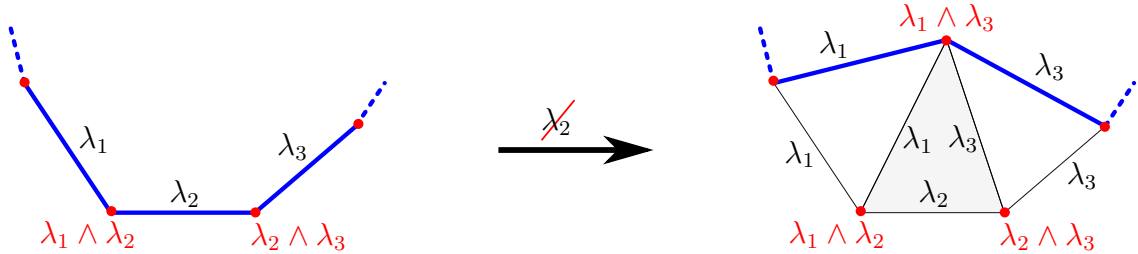
\caption{Triangulation of a null polygon Wilson loops into the two type of triangles.
The two types of triangles correspond to the ones bounded by $\lambda$'s which are all different or
the ones with the same $\lambda$ for all three edges.}\label{Triangulation}
\end{figure}

This is possibly interesting, because any null polygon can be decomposed into a sum of triangles of the two types.
Consider any three consecutive sides of a polygon, associated with twistors $\lambda_1$, $\lambda_2$, $\lambda_3$, and imagine replacing  the sequence of three sides
with two new sides, defined by the new vertex $\lambda_1 \wedge \lambda_3$. It is easy to see that this point is null-separated from the initial and final
vertices of the sequence of sides. The new polygon has the same sequence of twistors as the old one,
with $\lambda_2$ omitted (see figure \ref{Triangulation}). The difference between the old polygon and the new polygon is a pentagon. But the new vertex is actually null-separated from all points
in the old sequence of sides. The pentagon is really the sum of three null triangles, one of the first type (lying in the null plane $\lambda_1 \wedge \lambda_2 \wedge \lambda_3$)
and two of the second type (lying in the null planes $\lambda_1$ and $\lambda_3$).
Of course, it is possible to invert the roles of twistors and dual twistors in the above operation.

Iterating either operation, in any order, one can produce a variety of triangulations for any given polygon into triangles whose
vertices are a subset of all pairs of twistors, or all pairs of dual twistors in the polygon. Geometrically, these are
various intersections of the null-planes passing through the sides of the polygon.

Now, pick any such triangulation of two space-like separated polygons and write $r_{U(1)}(X,Y)$ as a sum over
pairs of triangles in the two triangulations. We can naturally decompose $r_{U(1)}(X,Y)$ in two pieces:
$r^+$ includes the contributions from triangles of the first type in $X$ paired up with triangles of the second type in $Y$, $r^-$
 includes the contributions from triangles of the second type in $X$ paired up with triangles of the first type in $Y$.

Notice that the symbols of $r^\pm$ contain only either type of twistorial cross-ratio, but have to add up to the symbol $S$ of
$r_{U(1)}(X,Y)$. Hence they must coincide with $S^+$ and $S^-$, and do not depend on the choice of triangulation.
The actual functions $r^\pm$ do not depend on the triangulation either: we can simply focus on one triangle in $X$, say of the first type,
and sum over the triangles in $Y$. This only contributes to $r^+$, but it is the full $r_{U(1)}$ between the triangle and
$Y$, hence it is independent of the triangulation of $Y$. Summing over triangles of the first type in $X$ we confirm that $r^+$ is independent of
the triangulation of $Y$. The same reasoning with triangles of the second type in $Y$ proves independence on the triangulation of $X$.

\subsection{Differential equations for $r^\pm(X,Y)$} \la{sec35}
In the OPE analysis, we will need to decompose $r_{U(1)}(X,Y)$ as a sum of contributions corresponding to the propagation of
a particle with given quantum numbers under various choices of conformal transformations.
In other words, we will need to decompose $r_{U(1)}(g \circ X,Y)$ in pieces which depend in specific ways on the
$SL(4)$ conformal transformation matrix $g$. We will particularly interested in the decomposition under a certain $SL(2)_\sigma \times SL(2)_\tau \times U(1)_\phi$
subgroup of $SL(4)$
\begin{equation}
g = \begin{pmatrix} g_\sigma \, e^{i \phi/2} & 0 \\ 0 &g_\tau \, e^{- i \phi/2} \end{pmatrix}
\end{equation}
These transformations preserve an $\mathbb{R}^{1,1}$ inside $R^{3,1}$: $U(1)_\phi$ rotations around the $\mathbb{R}^{1,1}$, $SL(2)_\sigma \times SL(2)_\tau$ conformal transformations in the $\mathbb{R}^{1,1}$
plane. Alternatively, they are the isometries of a $AdS_3 \otimes S^1$ spacetime, conformally equivalent to flat space.

When acting with $g$ on the $X$ or $Y$ polygons, we find that the two helicities $r^+$ and $r^-$ give rise to two towers of excitations, with a specific relation between the $\tau$ and $\sigma$ spins and the
$U(1)_\phi$ quantum number $m$. We can check directly the consequences on $r_{U(1)}(g \circ X,Y)$ of this fact. If we take $\lambda_i$ to be in the fundamental representation of $g$,
the Casimirs $C_\sigma$ and $C_\tau$ of $SL(2)_\sigma$ and $SL(2)_\tau$ satisfy
\begin{equation}
C_\sigma = \frac{m^2}{4} + \frac{m}{2} \qquad C_\tau = \frac{m^2}{4} - \frac{m}{2}\end{equation}
when acting on $r^+(X,Y)$, and
\begin{equation}
C_\sigma = \frac{m^2}{4} - \frac{m}{2} \qquad C_\tau = \frac{m^2}{4} + \frac{m}{2}\end{equation}
when acting on $r^-(X,Y)$. These Casimirs are given by $s(s-1)$ and we can call $s$ the spin. Note
that $s$ and $1-s$ give the same Casimir.

\section{Convolution with the anomalous dimension kernel}

\subsection{The anomalous dimension}
\label{anomdimsec}

In this section we recall the formula for the anomalous dimension of the operators that we
are exchanging.

As we discussed, in our problem it is convenient to focus on
 an $SL(2)_\tau \times SL(2)_\sigma \times SO(2)$ subgroup of $SL(4)$ \cite{OPEpaper}.
We consider a $U(1)$ gauge field and expand it into representations of these symmetries.
The states are multiplets of $SL(2)_\tau$ and all states in a given multiplet have the same
anomalous dimension \cite{Gaiotto:2010fk}. The dimension is the $L_0$ generator inside $SL(2)_\tau$.
The anomalous dimension depends on the momentum of the field. The momentum is a generator
inside $SL(2)_\sigma$. It turns out that the anomalous dimension depends also on the spin under
$SL(2)_\sigma$ of the field.

Thus, it is convenient to classify the states into primaries under $SL(2)_\tau$ and $SL(2 )_\sigma$.
It is convenient to think of the states as operators acting on the origin. In that language, the
$L_{-1}$ generators of $SL(2)_\tau$ is $\partial_-$ and the corresponding one in $SL(2)_\sigma$ is
$\partial_+$. The other two spacetime derivatives are just $\partial_z, \partial_{\bar z }$, and
carry charges $\pm 1$ under $SO(2)$.
The states are constructed from the field strength $F_{\mu\nu}$ and its derivatives. For the
purposes of classifying the states we can view the field strength as abelian and the derivatives
as ordinary partial derivatives. We also should impose the equations of motion and Bianchi
identities, $dF = d * F =0$. We are then interested in states which cannot be written
as $\partial_\pm$ of other states.
We can consider the twist operator which is the dimension minus the spin in the $+-$ plane, $\tau = \Delta - M_{+-}$, where $M_{+-}$
 is the spin in the $+-$ plane. We can define the conformal spin under $SL(2)_\tau$ as $\beta = {\tau \over 2 } $.   This
assigns zero twist to $\partial_-$. The twist corresponds to the
$\partial_\tau$ generator in $SL(2)_\tau$.
The states with lowest twist are the twist one states $F_{-z },~ F_{-\bar z}$ which have $SO(2)$ charges
plus or minus one. In this fashion we can classify all states and we get the following ones
\beqa \label{selfdf}
& &  F_{+-} - F_{z \bar z } ~,~~~~~~~~~~~ \partial_z^{m-1} F_{z - } ~,~{m\geq 1 }
~~~~~~~~~~~~~~~ \partial_{\bar z}^{- m-1} F_{\bar z +} ~,~~~m\leq -1
\\ \label{antiselfdf}
&&  F_{+-} + F_{z \bar z } ~,~~~~~~~~~~~ \partial_{\bar z}^{m-1} F_{\bar z - } ~,~{m\geq 1 }~~~~~~~~~~~~~~~ \partial_{ z}^{-m-1} F_{ z +}~,~~~m\leq -1
\\
 && s = \beta =1 ~,~~~~~~~~~~~~~~~ s-1 = \beta = { |m| \over 2 } ~,~~~~~~~~~~~~~ s= \beta -1 = { |m| \over 2} \label{confspins}
\eeqa
We have indicated the conformal spins under $SL(2)_\sigma$ and $SL(2)_\tau$ in the last line \nref{confspins}.
We see that we have two states at twist two, given by the first column. Then for each SO(2) charge
 $m$, $m\not = 0$,
  we have a states with twists $|m|$ or $|m|+2$ given by the second and third column respectively.
 The conformal spin $s$ of each state is given by  $ s = { \Delta + M_{+ -} \over 2 } $.

 The anomalous dimension of these states is then \cite{Basso}
\beq \label{forman}
 \gamma_{ 2 s} (p) =2g^2 \left[ \psi( s + i { p \over 2 } ) + \psi( s - i { p \over 2 } ) - 2 \psi(1) \right]
 \eeq
 where $\psi$ is Euler's psi function $\psi(x) = { \Gamma'(x )\over \Gamma(x) }$.
 Note that the first line \nref{selfdf}
 involves self dual fields and the second line  \nref{antiselfdf}
 involves anti-self dual fields.
 The conformal spin of the two towers is related to the $SO(2)$ charge.
 Thus, we can rewrite \nref{forman} in terms of the   $SO(2)$ charges as
 \beqa
{\rm For} ~~~~~ \nref{selfdf}: && \label{sdm} \gamma_{-{ m  } }( p) = \gamma_{2+ { m  }}(p)
 \\
{\rm For} ~~~~~  \nref{antiselfdf}: &&   \gamma_{  { m  }} ( p) = \gamma_{2-{ m  }}(p)
 \eeqa
Thus, we have now expressed the formulas for the anomalous dimensions in terms of the eigenvalues
 under some of the generators of the conformal group. Here $m$ denotes the $SO(2)$ charge, including
  its sign.
  Note that if $s$ is a half integer, then
 \nref{forman} is invariant under $s \to 1-s$, which implies that the anomalous dimensions
 depend only on the casimir of $SL(2)_\sigma$ (and the momentum, which is an element in $SL(2)_\sigma$).

 \subsection{ The anomalous dimension as a convolution }
\label{convsection}

 If we start from the one loop sources computed in the previous section and we Fourier transform them
 we can multiply each term by the corresponding anomalous dimension.

 Alternatively, we can transform a product of fourier transforms into a convolution.
 For that purpose it is useful to use the integral expression for the $\psi$ function
 \beq \label{integ}
 \psi(1 + z ) - \psi(1) = - 2 \int\limits_0^{\infty }\! dt\, {  e^{ - 2 z t } - 1  \over e^{2 t} -1}
  \eeq
  This integral expression is good for $z>-1$. Thus, in writing \nref{sdm} we have to separate the
  cases $m\geq0$ and $m<0$. We will get insertions of $e^{ \pm  t ( i P_\sigma \pm J )}$,
  where $P_\sigma$ is the momentum in the sigma direction and $J$ is the SO(2) charge, whose eigenvalue
  is $m$.
  When we insert this in the one loop source, it is the same as deforming the upper part of the polygon
  by an element of the conformal group.
  Using the integral expression we get that the convolution of the
  kernel with $r_+$  ends up giving us an expression of the form
  \beqa \label{convgam}
  \gamma * r_+  &=& - 4 g^2 \int\limits_{0}^{\infty}\!dt\, { \left[
   r_+( e^{ t ( i P - J )} X, Y) +  r_+( e^{ t ( - i P - J )} X, Y) - 2 r_+( X,Y)
   \right]_{\geq 0} \over e^{ 2 t} -1 }
 \\ && - 4 g^2 \int\limits_{0}^{\infty}\!dt\, {   \left[e^{ 2 t } \left\{
   r_+( e^{  { t   } ( i P + J )} X, Y) +  r_+( e^{  { t  } ( - i P + J )} X, Y) \right\} - 2 r_+( X,Y)
   \right]_{<0} \over e^{ 2 t} -1 } \nn
 \eeqa
 Where the symbols $[ \cdots ]_{\geq 0}$ and $[\cdots]_{<0}$ indicate  that we project
  on to the part that has positive (or zero)
 eigenvalues  (or frequencies) under $J$ or   negative eigenvalues  respectively.
 In performing these projections
 we assume that we are near the OPE limit, which gives a natural size to various terms, or a choice of contour.
  The last term, the one
 that is not shifted by $t$ can be dropped if one is computing the symbol, since it simply gives
 the one loop answer times a constant.

 These projection conditions can be done by performing contour integrals
 \beq \label{projec}
r_{\geq 0}( z )  = { 1 \over 2 \pi i } \oint\limits_{|w|>|z|} { d w \over ( w - z ) } r(w) ~,~~~~~~~~~
r_{<0}(z) = - { 1 \over 2 \pi i } \oint\limits_{|w|<|z|} { d w \over w -z } r (w)
\eeq
When we use these expressions in the integrand of \nref{convgam} we can view $e^{ \pm t J}$ as
acting on $z$. In other words, let us say that
 the original value of $z$, $z_0$ of the original polygon can
be viewed as a point on a circle, $z_0 = e^{ i \phi}$. Then in the positive frequency
 projector \nref{projec}, which acts on the  first line
of \nref{convgam},
we set $z = e^{- t } e^{ i \phi}$ and
$w = e^{ i \varphi + i \phi}$, and and integrate over $\varphi$. On the negative frequency projector
we simply have $z = e^t e^{i \phi}$ and the same $w$.
 In the end, we see that $r_+$ is evaluated at $z_0 e^{i \varphi}$.
This is the same as saying that the $X$ polygon is acted upon by $e^{ i \varphi J}$ where $J$
is the $SO(2)$ generator.

 We can combine the two integrals and write the full kernel acting on the positive helicity
 contribution as
  \beq \label{fkernel}
 \gamma * r_+=    4 g^2   \int\limits_{-\infty}^\infty dt \int\limits_0^{2 \pi} { d\varphi \over 2 \pi} \,
 { r_+( e^{   i { t  }  P + i \varphi J  } X, Y ) - r_+(X,Y) \over ( 1 - e^{ - t  } e^{  - i \varphi} ) ( 1 - e^{ t } e^{- i \varphi } ) }
 \eeq

 The kernel  for  $\gamma * r_-$ is the same except that
  we replace $e^{-i \varphi} \to e^{ i \varphi}$ in
 the kernel in \nref{fkernel}.
 Alternatively, we can use an expression for $\gamma * r_-$ which is similar to \nref{convgam} except
 that we exchange $J \to -J$ and $[\cdot]_{\geq 0} \to [\cdot]_{ \leq  0 } $, $[\cdot]_{<0 } \to [\cdot]_{> 0 } $.

 It is possible to do one of the integrals in a straightforward way (using mathematica, for example).
 The second integral then looks a bit ugly.
 (It can be done but it involves some square roots).
  Perhaps it becomes easier after simplifying
 various expressions.
 Ideally one would like to be able to do this integral at the level of symbols. We hope that
 this integral expression is of use in determining the symbol.
 If we perform the OPE expansion, by expanding in powers of $e^{ - \tau}$, then it is easy to
 do these integrals term by term since the projection onto positive or negative frequencies is trivial
 once we do the expansion in $e^{-\tau }$. Thus, this expression can be used to check that a
 possible guess does indeed obey the OPE expansion.

\section{The Hexagon Wilson loop}\la{hexagonsec}

A simple compact expression for the remainder function for
the hexagon Wilson loop was found in \cite{Volovich}, based on
previous work \cite{DelDuca:2009au,Anastasiou:2009kna}.
In this section we rederive the Hexagon Wilson Loop expectation value at two loops without using
these previous results.
 The main tools/assumptions will be that the OPE expansion holds and that the full result has a symbol. There are very strong self-consistency checks of both hypothesis so to a very great extent they are themselves proven.

The hexagon has three independent cross ratios $u_1,u_2$ and $u_3$ given by
\beq
u_i={x_{i-1,i+1}^2x_{i-2,i+2}^2\over x_{i-1,i+2}^2x_{i+1,i-2}^2}
\eeq
where the $x_i$'s are the cusps positions and $x_{i,j}^2=(x_i-x_j)^2$. We have $x_7=x_1$ etc.
It is also convenient sometimes to use the variables $b_1, b_2, b_3$ and $\mu$ given by
\beq
b_i=\sqrt{u_i\over u_{i-1}u_{i+1}}\ ,\qquad 
\mu= \Delta+\sqrt{\Delta^2-1}
\eeq
where $2\Delta=b_1b_2b_3-b_1-b_2-b_3$ and $u_4=u_1$ etc.
The two loop remainder function is a totally symmetric function of $u_1$, $u_2$ and $u_3$.
We can also express $\mu$ and $b_i$ in terms of momentum twistors\footnote{Here $\<\lambda_i,\lambda_j,\lambda_k,\lambda_l\>$ stands for $\lambda_i\wedge\lambda_j\wedge\lambda_k\wedge\lambda_l$.}
\beq
{ b_i \over  \mu^{ (-1)^i} } = \, { \langle \lambda_{ i-3} , \lambda_{ i-2} ,\lambda_{ i},\lambda_{ i+1} \rangle
\langle \lambda_{ i-4} , \lambda_{ i-3} ,\lambda_{ i-2},\lambda_{ i-1} \rangle
\over
\langle \lambda_{ i-3} , \lambda_{ i-2} ,\lambda_{ i-1},\lambda_{ i} \rangle
\langle \lambda_{ i-4} , \lambda_{ i-3} ,\lambda_{ i-2},\lambda_{ i+1} \rangle }
\eeq

There are three OPE expansion channels which correspond to the three inequivalent choices of two opposite edges along which we stretch the polygon. This is conformally equivalent to picking two adjacent legs which we make collinear. In each  OPE limit  one of the $u_i$'s goes to zero while the other two remain finite.
For example, in the channel where $u_2\to 0$ the OPE limit can be obtained by picking a reference square as indicated in \cite{OPEpaper} (see appendix F.3 in \cite{OPEpaper}).
 In terms of the symmetries of this reference square the three cross ratios read \cite{OPEpaper}\footnote{
 One minor change $\phi_{\rm here} = \phi_{\rm there} + \pi $. }
\beq
u_{1,3}=e^{\pm\sigma} { \sinh \tau \tanh \tau \over 2 ( \cosh \tau \cosh \sigma + \cos\phi ) } \ ,\qquad u_2 = { 1 \over \cosh^2  \tau}\ . \la{limitu2}
\eeq
The corresponding OPE limit is the limit where $\tau \to \infty $ with $\sigma$ and $\phi$ held fixed.

We will now describe the logic of the derivation of the two loop Hexagon remainder function. The technicalities of the several steps are then explained in detail in the subsections which will follow.
\begin{itemize}
\item We start by computing the $U(1)$ result in a given channel (\ref{roneu}), \nref{rtwou}.
All channels are equivalent so we can pick one of them, say the one where $u_2\to 0$, i.e. $\tau \to \infty$ with (\ref{limitu2}). We find\footnote{We use ${g_{YM}^2N_c\over16\pi^2}\equiv g^2 = a/2$ where $a$ is used in \cite{Bern:2005iz, Volovich}. This means that the two loop reminder function -- defined as what multiplies $g^4$ -- differs from the one in \cite{Volovich} -- defined as what multiplies $a^2$ -- by a factor of 4, $R_{\rm here}= 4 R_{\rm there}$.}
\beq
r_{U(1)}=-\text{Li}_2\left(1-u_1\right)+\text{Li}_2\left(u_2\right)-\text{Li}_2\left(1-
   u_3\right)+\log ^2\left(1-u_2\right)-\log \left(u_1\right) \log
   \left(u_3\right)+\frac{\pi ^2}{6} \la{Uone}
\eeq
This is the seed of the computation. It is computed in greater detail in section \ref{u1sec}.
\item
Next we rewrite $r_{U(1)}$ as a sum over particles propagating in the flux tube. At this order the particles are free and their energies are quantized to integers. The $SL(2)$ symmetry of the two null lines is preserved at this loop order and therefore the propagation of primaries and their descendants is conveniently packaged into $SL(2)$ conformal blocks. The relevant conformal blocks are computed in section \ref{blockssec} and read
\beq \la{confF}
\mathcal{F}_{\beta,p}(\tau)=\cosh^{-2\beta} (\tau)\ _2F_1\[\beta-i{p\over2}, \beta +i{p\over2},2\beta,\cosh^{-2}(\tau)\]\ ,
\eeq
where $p$ is the momentum conjugated to $\sigma$ and $\beta $ is the $SL(2)_\tau$
 conformal spin of the free particle states.
The set of primaries that can flow were described in section \nref{anomdimsec}.
\item
The seed (\ref{Uone}) is neatly decomposed in terms of exchanges of these particles as
\beq
r_{U(1)}=\int {dp\over2\pi}\, e^{-i p \sigma} \sum_{m=-\infty}^{\infty} \cos(m \phi)\, \mathcal{D}_{m}(p)\,{\cal F}_{{m}/2,p}(\tau)  \ ,\la{decompositionu1}
\eeq
where
\beq
\mathcal{D}_m(p)=4\,(-1)^{m} {B\(\frac{m}{2}+\frac{ip}{2},\frac{m}{2}-\frac{ip}{2}\)\over p^2+m^2}
\eeq
are the $U(1)$ form factors and $B(a,b)\equiv\frac{\Gamma(a)\Gamma(b)}{\Gamma(a+b)}$. This is a very non-trivial check of the OPE expansion. Formula (\ref{decompositionu1}) is derived in section \ref{decompositionsec} using a very useful differential operator
\beq
\Box =-\(\d_\phi^2+\d_\sigma^2\)\ .\la{boxop}
\eeq
It would be interesting to find a nice physical meaning for this operator.
\item Having identified the primaries that are flowing we can promote our result from one loop to two loops by correcting their free dimension to their quantum corrected anomalous dimension. More precisely
the two loop result when $u_2\to 0$ is given by
\beq
R_{2\, loops} = {1\over2} \log(u_2)\, D_2 + \widetilde R_2 \ ,
\eeq
where $D_2$ and $\widetilde R_2$ have regular power series expansions in $u_2 \propto e^{-2\tau}$.
$D_2$ is the discontinuity of $R_2$ around   $u_2 =0$.
We can identify $D_2$ by dressing the conformal blocks in (\ref{decompositionu1}) by the anomalous dimensions of the corresponding primaries   (\ref{forman}). That is\footnote{Note that $\gamma_{-m}(p)=\gamma_{m+2}(p)$.}
\beq
D_2=\int {dp\over2\pi}\, e^{-i p \sigma} \sum_{m=-\infty}^{\infty} \cos(m \phi)\, C_{m}(p)\,\gamma_{m+2}(p)\,\,{\cal F}_{{m}/2,p}(\tau)   \la{S2sum}
\eeq
In principle we could try to directly resum this expression, e.g. by making an educated ansatz for the kind of functions that could appear in the result. The result is given in formula (\ref{S2function}).
\item
A neater way of resumming (\ref{S2sum}) is by assuming that the resulting function has a symbol and computing this symbol. Then we find the function corresponding to this symbol and with the right reality properties and verify that it indeed resums (\ref{S2sum}). As argued in section \ref{S2symbolsec} the symbol of $D_2$ is given by
\beq
\text{Sym}\(D_2\)= u_1 \otimes X_1+u_3 \otimes X_3 + (1-u_2)\otimes Y_2  \la{S2symbol}\ .
\eeq
\item The functions $X_1$ and $X_3$ can be found by taking the limit where $u_1$ and $u_3$ are sent to  infinity respectively. They will be computed in section \ref{X1X3sec}.
\item The function $Y_2$ is related to the point where we reach the radius of convergence of the OPE expansion. That is the point where the   two null lines of the OPE expansion intersect. These
    are the two  null lines containing edges  $(x_3,x_4)$ and $(x_1, x_6)$, respectively.
     Once we know $X_1$ and $X_3$ we find $Y_2$ using the integrability condition which ensures that (\ref{S2symbol}) is the symbol of a function. This is done in section \ref{Y2sec} and concludes the re-sumation of (\ref{S2sum}).
\item Finally we need to find a totally symmetric function of $u_1,u_2$ and $u_3$ which behaves as
\beq
R_{2\, loops} =\frac{1}{2} \log(u_i) \,D_i  + \tilde R_i \qquad, \text{when $u_i\to0$}
\eeq
This is discussed in section \ref{symmetrizationsec}. It can be done in the most natural way using symbols. We assume that the two loops remainder function has a symbol. Then the simplest possible candidate is
\beq
\text{Sym}[R_{2\, loops}]=u_1 \otimes\text{Sym}[D_1] + u_2 \otimes\text{Sym}[D_2] + u_3 \otimes\text{Sym}[D_3] \la{guess}
\eeq
as explained in section \ref{OPEconstr}.
\end{itemize}

\subsection{The one loop source,  $r_{U(1)}$} \la{u1sec}

The seed for the perturbative OPE expansion in a given channel is the one loop $r_{U(1)}$ OPE
result (\ref{roneu}), \nref{rtwou}. As reviewed in section \ref{rev}, it is given by
\beq\la{ruone}
r_{U(1)}=\log\[{\<W\>\<W^{square}\>\over\<W^{top}\>\<W^{bot}\>}\]_{U(1)}\ .
\eeq
By computing that ratio explicitly, we find the result (\ref{Uone}). To be more precise, we find the result
\beqa
r_{U(1)}&=&-\text{Li}_2\left(1-u_1\right)+\text{Li}_2\left(u_2\right)-\text{Li}_2\left(1-
   u_3\right)+\log ^2\left(1-u_2\right)-\log \left(u_1\right) \log
   \left(u_3\right)+\frac{\pi ^2}{6}\nn\\&+&\log\(u_1/ u_3\)\log\(1-u_2\) \la{U1}
\eeqa
The expression (\ref{Uone}) is obtained from (\ref{U1}) by symmetrization under $u_1\leftrightarrow u_3$ which amounts to removing the last term $\log\(u_1/ u_3\)\log\(1-u_2\)$. Since that is a symmetry of the remainder function, such a symmetrization does not affect the final result.
 In fact, this last term is killed by the kernel that encodes the anomalous dimensions. The reason is
 that it corresponds to essentially zero momentum modes of the particle that has no anomalous dimension
 at zero momentum (as explained in \cite{OPEpaper} these can be viewed as Goldstone modes
 of some generators in $SL(2)_\tau$).
 Since the technical details of the computations are also relevant for the next section, we will now present them in some detail.

To compute $r_{U(1)}$, we start by making the following choice of momentum twistors
 for the reference square
\beq
\(\begin{array}{c}\lambda_\text{left}\\ \lambda_\text{top}\\ \lambda_\text{right}\\ \lambda_\text{bottom}\end{array}\)=\(\begin{array}{cccc}1&0&0&0\\0&0&0&1\\0&1&0&0\\0&0&1&0\end{array}\)\ . \la{square}
\eeq
To identify the left and right null lines of the square we construct the dual momentum twistors $\hat\lambda_{\rm left}=\lambda_{\rm bottom } \wedge \lambda_{\rm left} \wedge \lambda_{\rm top} =(0,1,0,0) $ and $\mu_{\rm right}=\lambda_{\rm top } \wedge \lambda_{\rm right} \wedge \lambda_{\rm bottom} =(1,0,0,0) $. The two null lines are
$\{\lambda_{\rm left},\mu_{\rm left} \}$ and $\{\lambda_{\rm right},\mu_{\rm right}\}$.

With this choice, the symmetries of the reference square (\ref{square}) and the $SL(2)$ symmetry of the two null lines act as multiplication by the $SL(4)$ matrices
\beq\la{trans}
U(s,t,f)=\(\begin{array}{cccc}e^{f/2+s}&0&0&0\\0&e^{f/2-s}&0&0\\0&0&e^{t-f/2}&0\\0&0&0&e^{-t-f/2}\end{array}\)\ ,\qquad
U_{SL(2,R)}=\(\begin{array}{cccc}1&0&0&0\\0&1&0&0\\0&0&a&b\\0&0&c&d\end{array}\)\ ,
\eeq
where $ad-bc=1$. Here, $t$ is associated with the ``Hamiltonian" generator that that moves points on the left and right sides of the reference square towards the bottom side, $s$ is associated with the ``Momentum" generator that that moves points on the bottom and top sides of the reference square towards the left side. The coordinate $f$ is associated with the $SO(2)$  ``Angular Momentum" which parametrize rotations (boosts) in the two transverse directions to the reference square. Up to a global $SL(4)$ transformations and rescaling, the Hexagon momentum twistors can be brought to the form
\beq\la{frame}
\(\begin{array}{c}\lambda_{\rm left}\\ \lambda_2\\ \lambda_3\\ \lambda_{\rm right}\\ \lambda_5\\ \lambda_6\end{array}\)=\(\begin{array}{cccc}1&0&0&0\\-1&0&0&1\\0&1&-1&1\\0&1&0&0\\0&e^{f/2-s}&e^{t-f/2}&0\\e^{f/2+s}&0&e^{t-f/2}&e^{-t-f/2}\end{array}\)\ .
\eeq
\begin{figure}[t]
\centering
\def\svgwidth{10cm}
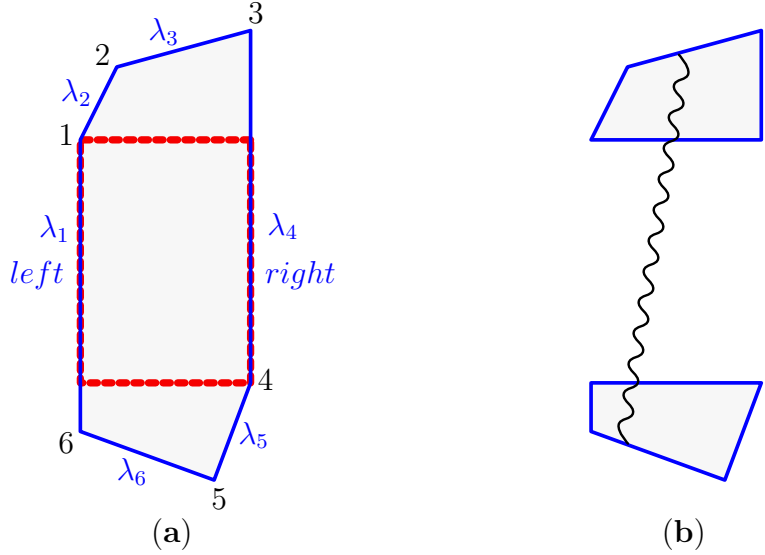
\caption{(a)  We see the hexagon. The sides 1 and 4 are the left and right sides.
The vertices 1 and 4 coincide with the reference square. In (b) we see the contours that give rise to the U(1) source $r_{U(1)}$. }\label{hexagonfigure}
\end{figure}
The corresponding top and bottom Pentagons are
\beqa
\text{Top}&=&(\lambda_{\rm left},\lambda_2,\lambda_3,\lambda_{\rm right},\lambda_{\rm bottom})\\ \text{Bottom}&=&(\lambda_{\rm left},\lambda_{\rm top},\lambda_{\rm right},\lambda_5,\lambda_6)\ .\nn
\eeqa
Indeed two of the null edges of all these polygons lay on the two null lines identified above. For example, the dual momentum twistor $\tilde\mu_{\rm left} =\lambda_6 \wedge \lambda_{\rm left} \wedge \lambda_{2} \propto \mu_{\rm left}$ and so on. Note also that the reference square (\ref{square}) has two cusps that coincide with two of the hexagon cusps. That is, $\lambda_{\rm right}\wedge\lambda_5=\lambda_{\rm right}\wedge\lambda_{\rm bottom}$ and $\lambda_{\rm left}\wedge\lambda_2=\lambda_{\rm left}\wedge\lambda_{\rm top}$.

This choice of reference square is convenient for $r_{U(1)}$ to be finite and
free of conformal anomalies and therefore to be a function of the conformal cross ratios $u_i$.  On the other hand, $\tau,\sigma$ and $\phi$ in (\ref{limitu2}) are associated with the three symmetries of a different reference square. That reference square was considered in \cite{OPEpaper} and realizes the $u_1\leftrightarrow u_3$ symmetry as $\sigma\leftrightarrow -\sigma$. The relation between the $t,s,f$ conformal frame and the $\tau,\sigma,\phi$ conformal frame is\footnote{Actually, in \cite{OPEpaper} we really had $f_{\rm here} = i \phi_{\rm there}  + i \pi$. Here we have a slightly different definition which introduces some irrelevant
minus signs.}
\beq
e^{t}=\sinh\tau\ , \qquad  e^{s}=e^{\sigma}\coth\tau
\ ,\qquad f= i \phi\
\eeq
The hexagon, pentagon and square polygon Wilson loop in a $U(1)$ theory can be found in \cite{Bern:2005iz}. Plugging these into (\ref{ruone}), using
\nref{distance}
and symmetrizing in $u_1\leftrightarrow u_3$, we arrive at (\ref{Uone}).

\subsection{Conformal blocks} \la{blockssec}

We interpret $r_{U(1)}$ as a sum of excitation freely propagating on the flux tube. These excitations are created at the bottom and been absorbed at the top, see figure \ref{hexagonfigure}(b).
When we will go from one to two loops, both the propagation as well as the creation and absorption of excitation are corrected. At one loop, the flux tube is invariant under the $SL(2)$ transformations $U_{SL(2,R)}$ which act on the bottom of the square. Therefore, the two loop correction to the propagation of an excitation due to interaction with the flux tube does not break  the $SL(2)$ symmetry
\cite{Gaiotto:2010fk}. It commutes with it. This means that the correction to the energy of excitation is the same for the primaries and descendants. This is why it so convenient to decompose the excitations of the flux tube into irreducible representations of $SL(2,R)$. The blocks of excitations forming an irreducible representation are functions of $t,s$ and $f$ known as \textit{conformal blocks}. Different blocks are uniquely identified (up to overall normalization) by the $SL(2)$ Casimir. To compute the hexagon conformal blocks, we first need to understand how $t,s$ and $f$ transforms when we act with the $SL(2)$ transformation in (\ref{trans}) on the bottom of the hexagon. We will then express the $SL(2)$ Casimir as a second order differential operator in $s,t$ and $f$.

Under the $SL(2)$ transformation $U_{SL(2,R)}$, $s,t$ and $f$ transform as
\beq\la{action}
e^{2t'}=(a+c)(b+a\,e^{2t})\ ,\qquad e^{s'+t'}=a\,e^{s+t}\ ,\qquad f'=f\ .
\eeq
Since $f$ does not transform, the conformal blocks only depend on $s$ and $t$. The transformation (\ref{action}) does not respect the naive group multiplication. The reason is the following. The cross ratios are $SL(4)$ invariant. Therefore, we can not ask how they transform under a specific $SL(2)$ group element without first fixing a frame. The hexagon parameterization (\ref{frame}) is one such choice of frame. After we act with $g_1\in SL(2)$ on the bottom, the frame changes. To rotate it back to the original frame, we act with an $M(g_1)\in SL(4)$ compensating transformation and rescale the twistors accordingly.\footnote{Explicitly we have
\beq
M(g_1)=   \sqrt{a\over a+c}\left(
\begin{array}{cccc}
1 & 0 & 0 & 0 \\
0 & \frac{a+c}{a} & 0 & 0 \\
0 & 0 & \frac{a+c}{a} & 0 \\
0 & 0 & -\frac{c}{a} & 1
\end{array}
\right)\ .
\eeq}
If we now act with a second $g_2\in SL(2)$ transformation on the bottom, the result is not the same as acting with $g_2g_1$ because $g_2$ do not commute with $M(g_1)$. Instead, we should act with $M(g_2g_1)\, g_2\, M^{-1}(g_1)$. In other words, when constructing the Casimir operator, we should then use the covariant derivatives.

There is a shortcut however. The $SL(2)$ Casimir can be written as
\beq\la{Casimir}
c_{SL(2)}=-J_1^2+J_2^2-J_3^3\ ,
\eeq
where for irreducible representation of spin $\beta$ we have
 $c_{SL(2)}=\beta(\beta-1)$. Since the finite action of any group element is given by (\ref{action}), we can read the action of $J_i^2$ directly by picking a group element $U_{SL(2,R)}=\mathbb{I}_{2\times 2} \times  e^{\theta\,J_i}$, expanding the result (\ref{action}) in $\theta$ and picking the quadratic term. We can then sum over $i=1,2,3$ to get the action of the Casimir operator. We find the corresponding differential equation for the conformal block\footnote{Note that contrary to the $\mathbb{R}^{1,1}$ case, where the $SL(2)$ conformal block is simply $F=\log(1+e^{-2t})$, here $s$ enters the differential equation.}
\beq
\[e^{-2t}\d_s^2+2(1-e^{-2t}\d_s)\d_t+(1+e^{-2t})\d_t^2-4\beta(\beta-1)\]F(s,t)=0\,. \nn
\eeq
It is very convenient to switch to the $\tau,\sigma,\phi$ variables in terms of which the $\sigma \to -\sigma$ symmetry is manifest. We find
\beq
 \[\d_\tau^2+2\coth(2\tau)\,\d_\tau-p^2{\rm sech}^2(\tau)\]e^{ip\sigma}\mathcal{F}_{\beta,p}(\tau)=4\beta( \beta-1)
 e^{ip\sigma}\mathcal{F}_{\beta,p}(\tau)\ . \la{difeq}
\eeq
where $p$ is the conjugate momentum to $\sigma$. The solution to this differential equation with the appropriate large $\tau$ behavior is given by (\ref{confF}). It is of course $p\leftrightarrow -p$ symmetric.
For future use let us quote the large $\tau$ behavior of these conformal blocks,
\beq
\mathcal{F}_{m/2,p}(\tau)=\frac{1}{\cosh^m\tau}\(1+ \frac{m^2+p^2}{4m}\frac{1}{ \cosh^2 \tau}+\dots \) \,. \la{largetau}
\eeq
We will now decompose the $U(1)$ contribution in terms of these conformal blocks.

\subsection{Decomposition of $r_{U(1)}$
 using $\Box\, r_{U(1)}$} \la{decompositionsec}

We will now decompose the $U(1)$ part using the conformal blocks (\ref{confF}) derived above. As explained in section \ref{anomdimsec} we expect two infinite towers of conformal primaries. In one of the towers the primaries have conformal spin $|m|/2$ and $SO(2)$ angular momentum $m$. In the other tower the conformal spin is $|m|/2+1$ for the same angular momentum $m$. We therefore expect, and will derive next, that $r_{U(1)}=r_++r_-$, where
\beqa
r_+=\int {dp\over2\pi}\, e^{-i p \sigma}\[\sum_{m=1}^{\infty}e^{im\phi}C_{m}(p){\cal F}_{{m}/2,p}(\tau)+\sum_{m=0}^{\infty}e^{-im\phi}\widetilde C_{m}(p){\cal F}_{{m}/2+1,p}(\tau)\] \la{decompu1}
\eeqa
and $r_-$ is obtained from $r_+$ by replacing $e^{i\phi}\to e^{-i\phi}$.
A simple way of proving this claim without decomposing the result is by using the projectors
\beq
{\cal D}_\pm\equiv \d_\tau^2+2\coth(2\tau)\,\d_\tau+{\rm sech}^2(\tau)\,\d_\sigma^2+\d_\phi(\d_\phi\mp2i)   \la{Dpm}
\eeq
which project out each of the two towers ${\cal D}_\pm r_\pm=0$.\footnote{Also ${\cal D}_\pm r_\mp=\mp4i\d_\phi\, r_\mp$. The building blocks $r_\pm$ should be identified with $r_+$ and $r_-$ discussed in section \ref{sec3}. See in particular the discussion in section \ref{sec35} which leads to this identification up to simple terms in the Kernels of the differential operators  ${\cal D}_\pm $.} Acting with these projectors directly on (\ref{Uone}) leads to
\beq
{\cal D}_+{\cal D}_-\, r_{U(1)}=0\
\eeq
which implies (\ref{decompu1}). Of course, after deriving $C_m(p)$ and $\widetilde C_m(p)$ we can trivially confirm directly that (\ref{decompu1}) holds.

To derive these structure constants it turns out to be quite useful to use the box operator defined in (\ref{boxop}). It is a sort of two dimensional Laplacian in the directions $\phi$ and $\sigma$ and turns out to simplify the $r_{U(1)}$ dramatically. In Fourier it amounts to multiplying the integrand in (\ref{decompu1}) by $(p^2+m^2)$. The reason for the remarkable simplification of $r_{U(1)}$ still eludes us. We find
\beq
\Box\, r_{U(1)}=\frac{-4}{1+{\rm sec} (\phi)\cosh(\sigma)\cosh(\tau)}  =4 \sum_{n=1}^{\infty} (-1)^{n}\( \frac{\cos(\phi)}{\cosh(\sigma)\cosh(\tau)}\)^n
\la{bR}
\eeq
The Fourier transform in $\sigma$ yields
\beqa
\Box\, r_{U(1)}= -\sum_{n=1}^{\infty} \int \frac{dp}{2\pi} e^{i p \sigma}\, (-2)^{n+1} B\(\frac{n}{2}+i\frac{p}{2},\frac{n}{2}-i\frac{p}{2}\) \( \frac{\cos(\phi)}{\cosh(\tau)} \)^n  \nn
\eeqa
For large $\tau$ we find therefore
\beqa
r_{U(1)}= 4\sum_{m=0}^{\infty}\int \frac{dp}{2\pi} e^{i p \sigma} \frac{\cos(m \phi)}{\cosh^{m}(\tau)}\,\frac{ (-1)^{ m}}{2^{\delta_{m,0}}} \frac{B\(\frac{m+ip}{2},\frac{m-ip}{2}\)}{(p^2+m^2)}   \[1+\frac{(2+m)(m^2+p^2)}{4  m(m+1)\cosh^2\tau} +\dots\] \nn
\eeqa
where we inserted back the denominator $(p^2+m^2)$ to remove the box.\footnote{At the end of the day we directly check the decomposition and thus verify that we did not loose any zero modes.}
Using (\ref{largetau}) and comparing this expansion with the expansion of (\ref{decompu1}) we read of the structure constants. We find
\beq
r_{U(1)} =\int {dp\over2\pi}\, e^{-i p \sigma}   \( \sum_{m=1}^{\infty}\, \frac{   \cos(m\phi)}{p^2+m^2} +\sum_{m=2}^{\infty}\,\frac{  \cos((m-2)\phi) }{p^2+(m-2)^2} \)\mathcal{C}_{m}(p)\,{\cal F}_{{m}/2,p}(\tau)  \la{decomp2}
\eeq
where
\beqa
\mathcal{C}_m(p)=4\,(-1)^{m} B\(\frac{m}{2}+\frac{ip}{2},\frac{m}{2}-\frac{ip}{2}\)
\eeqa
It is easy to see that the decomposition (\ref{decomp2}) is of the expected two towers form (\ref{decompu1}). It is also equivalent to the expansion (\ref{decompositionu1}) presented above. To show this one simply needs to use the identity $\mathcal{C}_{m}(p)\,{\cal F}_{{m}/2,p}(\tau) =\mathcal{C}_{2-m}(p)\,{\cal F}_{{2-m}/2,p}(\tau)$ whicg holds for integer $m$.

The structure constants $\mathcal{C}_m(p)$ are meromorphic functions which vanish very fast at infinity and which have infinitely many poles along the imaginary axis. We can compute the integral (\ref{decomp2}) by residues and hence obtain, for each $m$, an expansion in powers of $e^{-|\sigma|}$ as expected.

\subsection{Two Loops Resummation}

Now we will promote our one loop seed to two loops. When going from one two two loops three types of corrections. The particles energies acquire anomalous dimensions, the form factors are corrected and finally we have two particle exchange. The energies appear in the exponent multiplying $\tau \propto \log(u_2)$. Therefore, when we expand the result in powers of the coupling, the anomalous dimensions of the excitations, which appear in the exponent, will lead to a contribution to the remainder function which is linear in $\tau$. The other two effects do not contribute to the $\tau$ linear part. That is, at large $u_2$ we have
\beq
R_{2\, loops} = {1\over2} \log(u_2)\, D_2 + \widetilde R_2 \ ,
\eeq
where $D_2$ and $\tilde R_2$ have regular Taylor expansions in powers of $u_2$.
Using the OPE expansion we can compute $D_2$ by simply dressing the expansion (\ref{decomp2}) by the anomalous dimensions (\ref{forman})\nref{sdm}   of each of the conformal blocks,
\beq
D_2=\int {dp\over2\pi}\, e^{-i p \sigma}   \( \sum_{m=1}^{\infty}\, \frac{\gamma_{m+2}(p)  \cos(m\phi)}{p^2+m^2} +\sum_{m=2}^{\infty}\,\frac{\gamma_{m-2}(p) \cos((m-2)\phi) }{p^2+(m-2)^2} \)\mathcal{C}_{m}(p)\,{\cal F}_{{m}/2,p}(\tau) \,. \la{S2exp}
\eeq
The challenge is to perform this sum.

There is a simple brute force way of doing it if we assume that the result is an expression of transcendentality three given by a bunch of logarithms and polylogarithms. There are only a finite number of reasonable arguments which we can expect to encounter since we do not want to have discontinuities or other singularities at non-physical loci. Hence any such ansatz will have a very large but finite number of terms. We can then expand the ansatz at large $\tau$ and match it with the expansion of (\ref{S2exp}) to fix the coefficient of each of the logarithms and polylogarithms. This would be the analogue of the unitarity method approach where one makes an ansatz for the result using a basis of box integrals whose coefficients are fixed by matching with the expected discontinuities of the result.

We will follow another approach which uses the technology of symbols to break the computation of the sum (\ref{S2exp}) into much simpler blocks.

\subsubsection{The symbol of $D_2$}
 \la{S2symbolsec}

We assume that $D_2$ has a symbol. At the end of the day, when we will compute $D_2$ and compare directly with (\ref{S2exp}) we will check that this assumption is correct but for now let us motivate it
and understand its consequences.

Under the assumption that $D_2$ has a symbol, that symbol should be of the form (\ref{S2symbol})
\beq
\text{Sym}[D_2]=u_1\otimes \text{Sym}[X_1]+u_3 \otimes \text{Sym}[X_3]+(1-u_2)\otimes \text{Sym}[Y_2] \la{symbolS2}
\eeq
as we now argue.

First note that the function $D_2$ is the discontinuity of the two loops remainder function around $u_2=0$, where edges 1 and 2 become collinear\footnote{or, conformally equivalently, 4 and 5}. What singularities/discontinuaties can $D_2$ have? First, it cannot have another discontinuity at $u_2=0$. Such a second discontinuity would mean that $r_{U(1)}$ has a term linear in $\tau$ in contradiction to the OPE expansion. $D_2$ can have a discontinuity when two other edges become collinear. These are the points where edge 2 and edge 3 become collinear\footnote{or, conformally equivalently, 5 and 6} or when edge 3 and edge 4 become collinear\footnote{or, conformally equivalently, 1 and 6}. At these points $u_3\to0$ and $u_1\to0$ correspondingly.

The other singularity that $D_2$ may have is at the radius of convergence of the OPE expansion. That is the point where the   two null lines of the OPE expansion intersect. These are the two
 null lines containing edges $(x_3,x_4)$ and $(x_1, x_6)$, respectively.
  At that point $u_2=1$. This self-crossing does not happen in the Euclidean sheet.
These three possible singularities are precisely the ones captured by (\ref{symbolS2}).

The only other singularities/discontinuaties one may expect are when two cusps $x_i$ and $x_{i+3}$ become null separated. At these points, one of the $u_i$'s diverge. These are however already included in (\ref{symbolS2}).

We conclude that to derive (the symbol of) $D_2$ we can focus on finding (the symbol of) $X_1$, $X_3$ and $Y_2$. This is what we do in the next two subsections.

\subsubsection{The symbol of $X_i$} \la{X1X3sec}
In this section we focus on computing the contributions $X_1$ and $X_3$ to the sum (\ref{S2exp}). They are related by symmetry, $X_1=\left. X_3 \right|_{u_1 \leftrightarrow u_3}$, and therefore it is enough to
focus on $X_3$ which is what we will now do.

The contribution to $X_3$ comes from the large $\sigma$ behavior of $D_2$. More precisely, we should compute the $\sigma$ linear term of $D_2$ when $\sigma \to  \infty$.

Those come from \textit{double} poles in $p$. The residue of a double pole is the derivative of what multiplies the double pole. This derivative will, in particular, act on the exponential $e^{-i p \sigma}$ bringing down an $- i \sigma$ factor. We are not interested in the the action of the derivative on the rest since it will not lead to a contribution linear in $\sigma$. That is, we can simply replace the double poles by $-i \sigma$ times the factor which multiplies the double pole evaluated at the position of the double pole. Then we sum over all possible double poles. In this way we compute the term linear in $\log u_3$. The details of the summation are given in appendix \ref{X3appendix}. We find
\beqa
X_3&=&
-2 \text{Li}_2\left(-\frac{1}{\mu  b_1}\right)-2 \text{Li}_2\left(-\frac{\mu
   }{b_1}\right)-\text{Li}_2\left(1-u_1\right)-\text{Li}_2\left(1-u_2\right)-\text{Li}_2\left(1-u_3\right)\nn\\&&-\log
   ^2\left(u_1\right)-\log \left(1-u_2\right) \log \left(\frac{u_2}{u_1}\right)-\log \left(1-u_3\right) \log
   \left(\frac{\left(1-u_2\right) u_3}{u_1}\right)\ .\nn
\eeqa
$X_1$ is obtained from $X_3$ by $u_1\leftrightarrow u_3$, $b_1\leftrightarrow b_3$. At the level of symbols we have
\beqa
{1\over2}\text{Sym}[X_1]&=&2\, \frac{\mu  \left(b_3+\mu \right)}{b_3\, \mu +1}\otimes \mu -\frac{1-u_1}{u_3}\otimes
   \frac{1-u_2}{u_2}-\frac{1-u_2}{u_3}\otimes \frac{1-u_1}{u_1}+u_3\otimes \frac{1-u_3}{u_3} \nn\\
{1\over2}\text{Sym}[X_3]&=&2\, \frac{\mu  \left(b_1+\mu \right)}{b_1\, \mu +1}\otimes \mu -\frac{1-u_2}{u_1}\otimes
   \frac{1-u_3}{u_3}-\frac{1-u_3}{u_1}\otimes \frac{1-u_2}{u_2} +u_1\otimes \frac{1-u_1}{u_1}\nn
\eeqa

\subsubsection{The symbol of $Y_i$} \la{Y2sec}
The goal of this section is to constrain the remaining building block, $Y_2$. Since $Y_2$ is the discontinuity of the result at the boundary of the convergence region it is not straightforward to compute it using the expansion (\ref{S2exp}). Instead we will finding using some symbol technology.

To start we need an ansatz for $Y_2$. Looking at the symbols of $X_1$ and $X_3$ we see that in their last slot there is always either $\mu$ or $1-1/u_j$. Recall that derivatives of the symbol act on the last slot. Zeros and poles in the last slot translate into pole singularities for the derivative of the corresponding function and therefore have a physical meaning. The simplest ansatz for $Y_2$ which does not introduce new singularities is therefore
\beq
{1\over2}\text{Sym}[Y_2]= f_0 \otimes \mu + \sum_{i=1}^3 f_i \otimes \(1+1/u_j\) \,.
\eeq
We will see that this ansatz is indeed general enough. To determine $f_{0,1,2,3}$ we impose that $Y_2$ is a symbol of a function (\ref{testsymb}). That condition almost fix $Y_2$ completely. The details are given in appendix \ref{Y2appendix}. We find that if and only if
\beq
{1\over2}\text{Sym}[Y_2]=-u_1\otimes \frac{1-u_1}{u_1}-u_3\otimes \frac{1-u_3}{u_3}-4 \mu \otimes \mu+f(u_2)\otimes \frac{1-u_2}{u_2}
\eeq
then $\text{Sym}[Y_2]$ is the symbol of a function $Y_2$. The unfixed function $f_2(x)$ should be a rational function which we will now fix.

To find $f_2(x)$ we use the information that $D_2$ is given by a sum of the two towers of conformal blocks. In other words,
\beq
{\cal D}_{-}{\cal D}_{+} D_2=0 \la{last}
\eeq
where ${\cal D}_{\pm}$ project out each of the two towers, see (\ref{Dpm}).
The action of ${\cal D}_{-} {\cal D}_{+} D_2$ yields a term proportional to $\log(1-u_2)$ plus a term without any log. Both should be zero separately and hence we obtain two differential equations for $f_2$. One of them is
\beq
0=2x \,g''+g'-\frac{x^2-4x-1}{x(x-1)^2} \,, \qquad g=\log f_2 \la{wow}
\eeq
The other is a third order differential equation\footnote{It is given by $0=g'''+\frac{1-2x}{x(1-x)}g''+\frac{1-2x+3x^2}{x^3(x-1)^3}$}.
The solution of (\ref{wow}) is
\beq
f_2(x)= \frac{x}{(1-x)^2} \, C_2\,e^{-2 C_1/\sqrt{x}} \la{wow2}
\eeq
The constant $C_2$ is irrelevant for the symbol and we must set $C_1=0$ since $f_2$ should not be a transcendental function. Indeed, plugging (\ref{wow2}) in the above mentioned third order differential equation we see that we get zero iff $C_1=0$. Therefore, we obtain
\beq
{1\over2}   \text{Sym}[Y_2]=-u_1\otimes \frac{1-u_1}{u_1}-2 \left(1-u_2\right)\otimes \frac{1-u_2}{u_2}+u_2\otimes
   \frac{1-u_2}{u_2}-u_3\otimes \frac{1-u_3}{u_3}-4 \mu \otimes \mu\ . \nn
\eeq

\subsubsection{Final expression for the discontinuity $D_2$}

Once we have the symbol of the discontinuity $D_2$ it is easy to construct a function whose symbol is the same as the one of $D_2$. The function $D_2$
is then determined (up to a constant) by having only physical branch cuts \cite{Volovich}.
We find\footnote{The conformal cross ratios do not fix the polygon uniquely. As a result, the remainder function and, in particular, $D_2$ are multi-valued functions. Whenever we write such multi-valued function explicitly, we mean in the Euclidean sheet, namely, when all distances, that are not automatically null, are spacelike. }

\beqa
D_2&=&\left\{4 \left[\text{Li}_3\left(-\frac{1}{\mu
   b_1}\right)+\text{Li}_3\left(-\frac{\mu }{b_1}\right)\right]-2
   \left[\text{Li}_3\left(-\frac{b_2}{\mu }\right)+\text{Li}_3\left(-\mu
   b_2\right)\right]+\right.\nn \\
   &&2 \log \left(\frac{u_1}{u_3}\right)
   \left[\text{Li}_2\left(-\frac{1}{\mu
   b_1}\right)+\text{Li}_2\left(-\frac{\mu }{b_1}\right)\right]-\log
   \left(u_1 u_3\right) \left[\text{Li}_2\left(-\frac{b_2}{\mu
   }\right)+\text{Li}_2\left(-\mu  b_2\right)\right]-\nn \\&&
   \log \left(1-u_2\right)
   \left[2
   \text{Li}_2\left(1-\frac{1}{u_1}\right)+\text{Li}_2\left(u_2\right)+\frac{
   1}{3} \log ^2\left(1-u_2\right)+\frac{1}{2} \log
   ^2\left(\frac{u_1}{u_3}\right)+\frac{\pi ^2}{6}\right]+\nn \\&&
   2 \log
  \left. \left(\frac{1}{u_1}-1\right) \log \left(u_1\right) \log \left(u_3\right)\right\} + \Big\{ 1 \leftrightarrow 3  \Big \} \la{S2function}
\eeqa
It is indeed straightforward to check with \verb"Mathematica" that this function indeed resums (\ref{S2exp}).\footnote{This can done either numerically with very high precision or by expanding this functions to extremely high orders in $1/\cosh(\tau)$ and comparing with the explicit expansion (\ref{S2exp}).} Equation (\ref{S2function}) coincides indeed with the discontinuity at $u_2=0$ of    the result in \cite{DelDuca:2009au,Volovich}.

\subsection{Symmetrization and full result} \la{symmetrizationsec}

So far, we have computed $D_2$, e.i. the term linear in $\tau$ of the two loop remainder function at large $\tau$. In other words, we computed its discontinuity in $u_2$. The two loop remainder function we are interested in $R_2$, is a symmetric function of the conformal cross ratios $u_{1,2,3}$ (that is cyclic and parity invariant) hows OPE expansion at small $u_2$ is given by $D_2$ (\ref{S2function}). That information is not sufficient to uniquely determine $R_2$. The situation is analogous to trying to determine a four point correlation function at some loops level from its discontinuities. On the other hand, if we assume that $R_2$ has a symbol, then there is a natural candidate for its symbol
\beq\la{SymR2}
\text{Sym}[R_{2\, loops}]= u_1 \otimes\text{Sym}[D_1] + u_2 \otimes\text{Sym}[D_2] + u_3 \otimes\text{Sym}[D_3]\,.
\eeq
The structure of the symbol (\ref{SymR2}) is manifestly symmetric in the $u_i$'s and was motivated in section \ref{rev}. 

More explicitly, using (\ref{symbolS2}) we see that the structure of the result is quite neat, 
\beq\la{SymR2}
\text{Sym}[R_{2\, loops}]= \sum u_i \otimes u_j \otimes X_{ij} + \sum u_i \otimes (1-u_i) \otimes Y_i
\eeq
In the previous subsections we computed $X_{21}$ and $X_{23}$ (which we denoted as $X_1$ and $X_3$ respectively) and $Y_2$. All other $X_{ij}$ and $Y_{i}$ are related to those by trivial relabelings. The several OPE limits commute and this implies $X_{ij}=X_{ji}$ which is indeed a property that we can directly check from the solutions that we get. 

A simple but important self-consistency check is that (\ref{SymR2}) is a symbol of a function. A priory that is not the case for generic $D_i$.
To check that  (\ref{SymR2}) is a symbol of a function note first that since the functions $D_i$ (\ref{S2function}) do have a symbol (\ref{symbolS2}), it is enough to apply the check (\ref{testsymb}) to the first two slots of (\ref{SymR2}). Applied to the first two slots, the check (\ref{testsymb}) yields zero provided $X_{ij}=X_{ji}$. That is indeed the case and follows from the the fact that the different OPE limits commute as we just explained.

Again, up to a constant, we can find the function whose symbol is given by (\ref{SymR2}) by requiring the presence of physical branch cuts only \cite{Volovich}. The resulting two loops remainder function  is then given by \cite{DelDuca:2009au,Volovich}
\beqa
R_2&=&4\sum_{i=1}^3\left\{\[ \text{Li}_4\(-{b_i\over\mu}\)+ \text{Li}_4\(-\mu b_i\)\]-{1\over2}\text{Li}_4\(1-{1\over u_i}\) -\log(b_i)\[\text{Li}_3\(-{b_i\over\mu}\)+\text{Li}_3\(-\mu b_i\)\]\right.\nn\\&&+{\log^2(b_i)\over2}\[\text{Li}_2\(-{b_i\over\mu}\)+\text{Li}_2\(-\mu b_i\)\] + {\log^3(b_i)\over12}\[\log\({(\mu +b_i)^2\over\mu b_i}\)+\log\({(1+\mu b_i)^2\over\mu b_i}\)\] \nn\\&&\left.+{1\over24}\[\log^4(\mu)+4\log^2(b_i)+2\pi^2\log^2(\mu)+{7\pi^4\over15}\]\right\}+{\pi^4\over18}\nn\\&&- { 1 \over  2}  \[\sum_{i=1}^3\text{Li}_2 \( 1 -{1\over u_i}\)\]^2+{\pi^2\over3}\[\sum_{i=1}^3\log\({1+\mu b_i\over\mu +b_i}\)\]^2+{1\over6}\[\sum_{i=1}^3\log\({1+\mu b_i\over\mu +b_i}\)\]^4
\eeqa

\section{Conclusions, discussion and speculations}

In this paper we have computed the discontinuities of the two loop remainder function
for polygonal null Wilson loops for a general number
of sides.
Here we considered the ``OPE discontinuities". These are defined as follows. 
We pick two sides of the null polygon, which divides the polygon in two halves. 
 We act on one half  of the
polygon with the operator $e^{ - \tau H}$ where $H$ is a ``Hamiltonian'' in SL(4) which preserves the
two null sides, then this Hamiltonian squeezes this half of the polygon. We can perform the
full OPE expansion of the one loop answer, keeping all powers of $e^{-\tau }$. This is what we call
the one loop result or $r_{U(1)}$, since it can be computed in a free $U(1)$ theory. The OPE expansion
is defined for any loop order. So for any loop order we can take a contour in the space of cross ratios
where we deform the contour by setting a large value of $\tau$, then we take $\tau \to \tau + 2 \pi i$.
In a perturbative expansion, this picks out all the logarithmic terms that arise due to anomalous dimensions.   We call this ``the OPE
discontinuity''. In the case of the two loop answer, there is only a linear term in $\tau$.
The coefficient of this linear term in $\tau$ is the discontinuity $D$ of the two loop answer. It is
a transcendentality three function. It can be computed by considering the one loop OPE and then
multiplying each term by the anomalous dimension. Since the anomalous dimension depends continuously
on a ``momentum'' quantum number, we find a final expression for $D$ which is has the form of a
convolution of a simple kernel acting on the one loop expression.
The final answer, given in \nref{convgam}\nref{fkernel}, is relatively simple and compact.
Note that in this form, it is quite manifest that the final answer is a transcendentality three function.
The one loop answer is a transcendentality two function and the integral \nref{convgam} raises the
transcendentality by one. The integral that projects onto positive or negative frequencies \nref{projec}
does not
raise the transcendentality because it is a closed contour integral.
Note that the transcendentality of the answer is not manifest in any of the previously known integral
representations for the Wilson loop or the amplitude.

It would be nice to be able to evaluate the symbol of \nref{convgam} or \nref{fkernel} without having
to evaluate the integrals. A direct evaluation of the integrals seems possible and doable (as long
as one has enough patience with mathematica...). On a term by term basis, these integrals seem to
give square roots of momentum twistor cross ratios. We do not know if they all cancel in the final
answer, since we have not computed all the terms.

We have argued that the knowledge of all OPE discontinuities should be enough to reconstruct 
the symbol of the two loop answer, for any number of sides. 

There are some terms in the symbol of the final two loop answer which we can predict in a simple way.
Namely, we can have terms of the form $u_{ij} \otimes u_{kl} \otimes X_{ij,kl} + u_{ij} \otimes
(1-u_{ij}) \otimes Y_{ij} $ where $i,k,j,l$ label sides of the polygon which  are cyclically ordered.
Here $u_{ij}$ are the cross ratios made with the four points at the ends of the two null sides.
These terms arise from ``crossed'' propagators. These are terms which appear in the remainder function
due to an oversubtraction when we say that $W = e^{ \Gamma_{cusp} A_{1-loop}  + R } $. These are terms
that are present in the exponential of the one loop answer but are not present in the {\it planar}
two loop answer. Thus they contribute to the remainder function. For some such terms, we can even say
what $X_{ij,kl}$ and $Y_{ij}$ are. If $i,j,k,l$ are well separated from each other,  these crossed
propagators are the only contribution to the symbol. Now, in the case of the hexagon, there are
other Feynamn diagram contributions. However, we have experimentally seen that the symbol of the
two loop answer still has only these two type of terms in the first slot. Some of us want to conjecture
that this is true in general. Other of us want to be more cautious since the evaluation of individual
terms in \nref{fkernel} give terms which are not of this form (which could eventually cancel).
Clearly, if one makes a correct hypothesis regarding the possible terms appearing in the symbol, then
the computation of the answer reduces to a finite problem, determining the precise combination of a
finite number of possible symbols. In that case, the explicit OPE discontinuities
we have discussed could give a large number of constraints which would help in determining the answer.

We have shown how to derive the answer for the hexagon using our method.
In fact, in the hexagon method we did not follow word by word the general route outlined
in the first couple of sections,  which we could have also done.  Instead,
 we
performed some explicit Fourier transformations, which lead to particularly simple formulas.
In particular, we noticed that the action of a particular differential operator ($ \Box$), simplified
the expressions considerably. It would be nice to see if this simplification
extends to other cases or whether it is just 
an accident for the hexagon. It is  reminiscent of the recursion relations for integrals
studied in \cite{Drummond:2010cz}. In the hexagon case, one could fix the discontinuity by looking
at its own discontinuities. Of course, one would like to be able to do this for the general case, so
as to compute the symbol step by step.

We also considered the heptagon Wilson loop (in appendix \ref{hybridap}). We have derived its OPE discontinuity in a particular mixed kinematical regime. That result can be used in the future to restrict the form of the heptagon two loops reminder function.

As we argued in \cite{Gaiotto:2010fk}, we expect our two loops results to be valid for
null polygon Wilson loops in any conformal gauge theory with a weak coupling limit.

Our expression for the discontinuity amounts to an insertion of the
anomalous dimension. In fact, we could imagine we inserting $e^{ 2 \pi i H } -1$, where $H$ is the full
Hamiltonian of the interacting theory. This gives us the discontinuity of the answer. It also has
the interpretation of adding (a combination) of conserved charges along the worldsheet (or planar
diagram). From the point of view of the amplitude, these are non-trivial integrability charges.
It should be interesting to understand how to add  more general integrability charges in
order to further constrain the problem.

In the particular case of three loops, we can find some of the discontinuities in the same way. 
For example, in the OPE limit, there will be $\tau^2$ terms which are given by applying the 
anomalous dimension  
kernel twice to the one loop answer. This constrains some of the terms in the symbol. 
In particular, we can determine terms like $ X_i .X_{i+2} \otimes X_{i}.X_{i+2} \otimes S $, by 
considering the OPE that leads to a collinear limit. 
In addition, we can do the following. Imagining that one had solved the two loop problem completely. 
Then one could find the non-logarithmic parts of the two loop OPE. This involves the exchange 
of two particle states. Then one can multiply these by the one loop anomalous dimensions and, together
with the two loop anomalous dimension acting on the one loop term, we can 
get the linear in $\tau$ terms in the OPE.

\subsection*{Acknowledgments}

We thank F. Alday, B. Basso,  F. Cachazo, S. Goncharov, J. Penedones, M. Spradlin, D. Skinner,  and A. Volovich  for discussions. This work was supported in part by   U.S.~Department of Energy grant \#DE-FG02-90ER40542. Research at the Perimeter Institute is supported in part by the Government of Canada through NSERC and by the Province of Ontario through MRI. D.G. is supported in part by the Roger Dashen membership in the Institute for Advanced Study. D.G. is supported in part by the NSF grant PHY-0503584. The research of  A.S. and P.V. has been supported in part by the Province of Ontario through ERA grant ER 06-02-293. Research at the Perimeter Institute is supported in part by the Government of Canada through NSERC and by the Province of Ontario through MRI. This work was partially funded by the research grants PTDC/FIS/099293/2008 and CERN/FP/109306/2009.

\appendix

\section{Simple examples}

In these section we will demonstrate by explicit examples how our techniques  can be used for obtaining new results at two loops. We will consider two examples: The first is a double scaling limit of the hexagon where a single primary per $SO(2)$ angular momentum propagates in the flux tube. The other is the full OPE discontinuity of the heptagon two loops remainder function in a mixed kinematical regime. Both examples can easily be extended to any number of edges.

\subsection{Double scaling limit}\la{Doublescaling}

In this section we will consider an interesting kinematical limit where the two loop computation simplifies dramatically. Remarkably the final form of the result in this limit\footnote{or rather of a particular symmetrization of the result in this limit} captures most of the features of the full result. The regime we want to consider is that when $i\phi,\tau \to +\infty$ diverge with $\zeta \equiv \tau-i\phi$ held fixed and large. In this limit $u_1$ and $u_3$ are finite while $u_2\to0$ so that the $U(1)$ result for the hexagon becomes
\beq
\hat r_{U(1)}(\sigma,\zeta)=g^2\[ \frac{\pi^2}{6}-\log(u_1)\log(u_3)-{\rm Li}_2(1-u_1)-{\rm Li}_2(1-u_3) \] \la{u1scaling}
\eeq
In this double scaling limit the second tower (i.e. sum) in (\ref{decompu1}) and (\ref{decomp2}) can be dropped and in the first sum we keep only the terms with positive $SO(2)$ charge. We can replace the conformal block by its leading large $\tau$ asymptotics. For example,
\beq
 \hat r_{U(1)}(\sigma,\zeta) =\int {dp\over2\pi}\,   e^{i p \sigma} \sum_{m=1}^{\infty} 2^{m-1}\frac{ \mathcal{C}_{m}(p)}{p^2+m^2} e^{-i p \sigma-m \zeta }    \la{decompdc}
\eeq
Physically, in that limit only primaries survive and, furthermore, only the primaries of one of the towers. That is, for each $SO(2)$ angular momentum we have a single primary propagating.

Next we want to promote this result to two loops by multiplying each summand by $\gamma_{m+2}(p)$. In this way we compute the double scaling limit of $D_2$. Alternatively, in the $\sigma$-space we can apply the anomalous dimension kernel as a convolution as described in section \ref{convsection}. In the scaling limit only one tower with $e^{i m \phi}$ with $m>0$ is propagating. Therefore we do not need to project the result into positive and negative frequencies and the result of applying the anomalous dimension kernel is very simple
\beq
 \hat D_{2}(\sigma,\zeta) = 4 g^2 \int\limits_{0}^{\infty} \frac{dt}{e^{2t}-1} \[ 2\,  \hat r_{U(1)}\(\sigma,\zeta\) - \hat r_{U(1)}(\sigma-t,\zeta+t) - \hat r_{U(1)}(\sigma+t,\zeta+t) \]
\eeq
It is straightforward to plug (\ref{u1scaling}) inside this integral and perform the integral using \verb"Mathematica" for example.\footnote{The kind of integrals we need to do are
\beq
\int_{1}^\infty \frac{dw}{w(w-1)} \log(a w+b)\log(c w+d) \qquad \text{and} \qquad \int_{1}^\infty \frac{dw}{w(w-1)} {\rm Li}_2\(1+\frac{a}{b+w}\)
\eeq
which are quite simple to handle.
} We find
\beqa \label{doublescal}
 \hat D_{2}(\sigma,\zeta)&=&4\[ \text{Li}_3\left(-\frac{1}{{\hat b}_1}\right)-
   \text{Li}_3\left(-{\hat b}_2\right)+ \text{Li}_3\left(-\frac{1}{{\hat b}_3}\right)\]-2\text{Li}_2\left(-{\hat b}_2\right) \log \left(u_1u_3\right)\\ &+&2\[\text{Li}_2\left(-\frac{1}{{\hat b}_1}\right)-\text{Li}_2\left(-\frac{1}{{\hat b}_
   3}\right)\] \log \left(\frac{{u}_1}{{u}_3}\right)
   + 2 \log [(1 + { 1 \over \hat b_1} )(1 + { 1 \over \hat b_3 } )] \log u_1 \log u_3  \nn
\eeqa
where $\hat b_1=b_1/\mu$, $\hat b_3=b_3/\mu$ and $\hat b_2=b_2 \mu$. We also used the equation
$ ( 1 + \hat b_2 ) = (1 + 1/\hat b_1 )(1 + 1/\hat b_3 )$, which is true in this limit.
The double scaling limit is a neat example when the one loop result can be promoted to a two loop prediction by means of a single and very simple integral which increases the transcendentalities of the functions in the $U(1)$ result by one unit.

It is curious to note that the double scaling result roughly captures half of the full result. It would be interesting to understand if it can be promoted in a simple way. This might be a shortcut towards the full result which might avoid the re-summations described in the main text.
More explicitly, note that if we start from \nref{doublescal}, add the same but with $\mu \to 1/\mu$, we
are  missing the third line in \nref{S2function}.

The scaling limit can be easily generalized to other polygons with more edges hence providing very non-trivial constraints on the form of the discontinuity of two loop MHV amplitudes for more than six gluons. A simple way to state the generalization is to say that we choose a special polygon where
all the momentum cotwistors (say $\hat \lambda_i$) of the top part of the polygon are the same and,
in addition, they
are orthogonal to $\lambda_L$, $\lambda_R$ which are the momentum twistors of the two special lines
that we are using to extract the OPE discontinuity.
Note that it makes sense to take this special polygon
only after extracting the original OPE discontinuity (the linear term in $\tau$ in the OPE).
For such polygons, only $r^+$ is nonvanishing, and
$r^+$ only has positive frequencies. Thus we can simply
 do the integral in the first line of \nref{convgam} to get the discontinuity.

\subsection{The Hybrid}\la{hybridap}

In this second example we consider the OPE discontinuity of a family of polygons we call \textit{Hybrid}'s. An Hybrid is a polygon where the top part is in $\mathbb{R}^{1,1}$ kinematics while the bottom is in general $\mathbb{R}^{1,3}$ kinematics. The physical reason for a big simplification in that regime is that the top part can only absorb $F_{+-}$ type excitation and therefore we will not have two infinite towers of conformal blocks. Still, the result is not trivial and, in particular, involve polylogarithms and not just logarithms as in strict $\mathbb{R}^{1,1}$ kinematics \cite{DelDuca:2010zp,Heslop:2010kq,Gaiotto:2010fk}.

\begin{figure}[t]
\centering
\def\svgwidth{4.5cm}
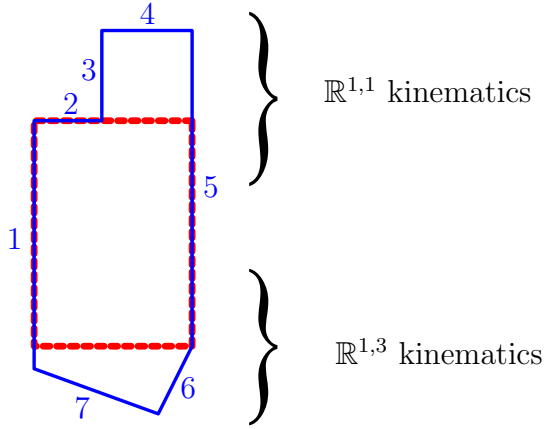
\caption{A polygon in a mixed kinematics called {\it Hybrid}. In the OPE limit $u\to\infty$ the two loops remainder function can be divided into two pieces as $R_2=t\,D+\widetilde D $. The function $D(s,t)$ is computed in this section.}\label{Hybrid}
\end{figure}

The simplest possible Hybrid is an heptagon where the top part (in $\mathbb{R}^{1,1}$) has three edges while the bottom part (in $\mathbb{R}^{1,3}$) has two edges and is the same as the bottom part of the hexagon (see figure \ref{Hybrid}).

This Hybrid has only two independent cross ratios. One way to see that is as follows. Given an heptagon hybrid, we can use conformal transformations to fix all points on the top as well as the bottom point which is shared by the reference square. To fix the remaining bottom point which is on one of the two null lines (the cusp $\lambda_7\wedge\lambda_1$) we need one parameter. We are left with the point \textit{outside} the plane, in the bottom (that is, the cusp $\lambda_6\wedge\lambda_7$). It is null separated from the two neighboring cusps and therefore depend on two more parameters. However we can use $SO(2)$ orthogonal transformations -- which leave all other points invariant -- to kill one of these parameters. Hence we have $1+1=2$ independent cross ratios. These two cross ratios read
\beq
y \equiv e^{2s}=\frac{\langle\lambda_1,\lambda_6,\lambda_2,\lambda_3\rangle  \langle\lambda_2,\lambda_7,\lambda_5,\lambda_6\rangle
   }{\langle\lambda_5,\lambda_6,\lambda_2,\lambda_3\rangle  \langle\lambda_6,\lambda_7,\lambda_1,\lambda_2\rangle } \,, \qquad \eta\equiv e^{2t}=\frac{\langle\lambda_1,\lambda_6,\lambda_4,\lambda_5\rangle  \langle\lambda_1,\lambda_2,\lambda_5,\lambda_7\rangle
   }{\langle\lambda_7,\lambda_1,\lambda_5,\lambda_6\rangle  \langle\lambda_1,\lambda_2,\lambda_4,\lambda_5\rangle }
\eeq
The two loop remainder function is given by
\beq
R_2(s,t)= t \,D(s,t)+ \widetilde D(s,t)
\eeq
where both $D$ and $\widetilde D$ have regular expansions in $1/\eta=e^{-2t}$. In this section we predict the function $D(s,t)$.
The large $t$ limit corresponds to the channel in figure \ref{Hybrid} . In this channel the  $r_{U(1)}$ is
\beq
r_{U(1)}=2 \,{\rm Li}_2(-1/x)-2 \,{\rm Li}_2(-1/y) \la{ru1hybrid}
\eeq
where
\beq
x=\frac{y \eta}{\eta-1} \equiv e^{2u}\ .
\eeq
We have\footnote{Integration slightly below the real axis.}
\beq \la{decomposition2}
r_{U(1)}={1\over2} \int dp \, \frac{e^{-i p u}-e^{-i p s}}{i \(\frac{p}{2}\)^2 \sinh\( \frac{\pi p}{2}\)}
\eeq
Note that the role of the second term in (\ref{ru1hybrid}) is minimal in the OPE expansion: When $\eta\to \infty$ we have $x\to y$ and therefore the second term is only canceling the $\eta$ independent constant coming from the OPE expansion of the first term. Inside (\ref{decomposition2}) we recognize the $F_{+-}$ form factor \cite{Gaiotto:2010fk}.

We will now argue that the term linear in $t$ of the two loop result has the remarkably simple form
\beq
D=-F(x)+F(y)\la{rhybrid}
\eeq
where
\beqa
F(x)&=&\text{Li}_3\(-x\)+2 \,\text{Li}_3\(\frac{1}{x+1}\)-
   \text{Li}_2\(-x\) \log\(\frac{x}{(x+1)^2}\)\\&+&{1\over3}\log
\(\frac{x^3}{x+1}\) \log ^2(x+1)+{\pi^2\over3} \log (x+1) \nn
\eeqa
As before the result (\ref{rhybrid}) is given by a difference of a function of $x$ and a function of $y$. As in the $U(1)$ case, the function of $y$ is simply regulating the result as $\eta\to \infty$ and therefore $x\to y$.

To learn what type of excitation are propagating, we act on $r_{U(1)}$ with the $SL(2)$ Casimir operator. The corresponding differential equation reads
\beq
{\cal D} \mathcal{F}_{\beta}(\eta,y) \equiv \[\eta(\eta-1) \partial_{\eta}^2+(2\eta-1)\partial_{\eta}+y \partial_{\eta} \partial_y \] \mathcal{F}_{\beta}(\eta,y)=\beta(\beta-1) \mathcal{F}_{\beta}(\eta,y) \la{dif}
\eeq
If we act with the differential operator ${\cal D}$ on the first term in (\ref{ru1hybrid}) we get zero. This is consistent with our physical picture: It means that (\ref{rhybrid}) is describing the exchange of a single primary with $\beta=1$ plus its descendants. This primary is nothing by $F_{+-}$! As argued above, the second term in (\ref{ru1hybrid}) is simply regulating the OPE expansion and therefore we will not care about it; in particular it does not carry any $\eta$ dependence.

Having identified the primary which is flowing we can now promote the result to two loops. We simply need to multiply the Fourier decomposition (\ref{decomposition2}) by $\gamma_2(p)$ (\ref{forman}) and perform the integration. Equivalently, we can use the convolution kernel for $\gamma_2(p)$. Doing so, we find (\ref{rhybrid}). More precisely, the OPE expansion tells us to dress the first exponent in (\ref{decomposition2}) by $\gamma_2(p)$; this leads to $F(x)$. We can think of $-F(y)$ as regulating the OPE limit $\eta\to \infty$ which leads to $x\to y$.

The computation of the full remainder function for the Hybrid does not seem, however, to be considerably simpler than the computation of the general heptagon. The reason is simple; the OPE expansion in other channels will take us out of the hybrid kinematics. In those channels we will have all primaries propagating and not just $F_{+-}$.

\section{Technical details for section \ref{hexagonsec}}

\subsection{More details on $X_3$}\la{X3appendix}

In this section we compute the discontinuity of $D_2$ around $u_3=0$. That is the function $X_3$ defined as the $\sigma$ linear term of $D_2$ when $\sigma \to + \infty$ (\ref{symbolS2}). This term comes from \textit{double} poles in $p$ of the Fourier transform of $D_2$ (\ref{S2exp}). The residue of a double pole is the derivative of what multiplies the double pole. This derivative will in particular act on the exponential $e^{-i p \sigma}$ bringing down an $- i \sigma$ factor. We are not interested in the the action of the derivative on the rest since it will not lead to a contribution linear in $\sigma$. That is,  we can simply replace the double poles by $-i \sigma$ times the factor which multiplies the double pole evaluated at the position of the double pole. Then we sum over all possible double poles. In this way we compute the term linear in $\sigma$.

For each $m$ there are infinitely many poles. More precisely, for each $m$,  we have the simple poles in the structure constants $\mathcal{C}_m(p)$ -- which are already present in the $r_{U(1)}$ result -- multiplying the simple poles in the anomalous dimensions $\gamma_m(p)$
\beq
\gamma_{m}(p) \simeq \frac{{ 4 i}}{p-i(m+2k)} \qquad \forall \qquad k\ge 0  \la{expgamma}
\eeq
These give rise to infinitely many double poles at integer imaginary values. These are not all double poles but almost. The remaining ones are very simple and finite (for each $m$) and will be identified and discussed below.

The key observation is that the poles of the anomalous dimension always have the same residue since they are made of polygamma function which are log derivatives. As seen in (\ref{expgamma}) the residue is $4i$. Hence, from the discussion above we conclude that the effect of the $\gamma_m(p)$ functions is trivial and simply gives  $4 i$ times $-i \sigma$ times the result without the gamma functions which is nothing but the $U(1)$ result! This is
\beq
(-2)X_3=-4\,r_{U(1)} + \[\begin{array}{ll}\text{contribution from finitely many}\\ \text{poles which we will now discuss}\end{array}\] \ .\la{X3structure}
\eeq
Let us now turn to the poles which we did not consider. They come from two places:

\begin{enumerate}
\item The pole at $p=im$ for $m\ge 1$ in the first tower in (\ref{S2exp})
\beq
\text{First tower}=\int {dp\over2\pi}\, e^{-i p \sigma}    \sum_{m=1}^{\infty}\, \frac{\gamma_{m+2}(p)  \cos(m\phi)}{p^2+m^2}  \mathcal{C}_{m}(p)\,{\cal F}_{{m}/2,p}(\tau)\ .\la{firsttower}
\eeq
The simple pole of $\mathcal{C}_{m}(p)$ at $p=im$ is not dressed by a pole of $\gamma_{m+2}(p)$, so we overshoot when writing $r_{U(1)}$ in (\ref{X3structure}) since $r_{U(1)}$ contains all poles. On the other hand $1/(p^2+m^2)$ does have a simple pole and therefore there is indeed a double pole at $p=im$ in (\ref{firsttower}). So we need to sum the contribution from the double pole from $1/(p^2+m^2)$ and subtract the $U(1)$ contribution from the double pole at $p=im$. The resulting contribution is
\beq
\mathcal{E}_1=8 \text{Li}_2\left(-\frac{1}{b_1 \mu }\right)+8 \text{Li}_2\left(-\frac{\mu }{b_1}\right)+4\log
   \left(\frac{\left(u_2-1\right) \left(u_3-1\right)}{u_1}\right) \log \left(-\frac{\left(u_2-1\right)
   u_3}{u_1}\right)\nn
\eeq
and should be subtracted in $X_3$ (\ref{X3structure}).

\item The double pole at $p=0$ for $m=2$ in the second tower of $D_2$ (\ref{S2exp}),
\beq
\left.\text{Second tower}\right|_{m=2}=\pint {dp\over2\pi}\, e^{-i p \sigma} \,\frac{\gamma_{0}(p)  }{p^2 } \,\mathcal{C}_{2}(p)\,{\cal F}_{1,p}(\tau) \ .
\eeq
Here, we have a double pole at $p=0$ in $r_{U(1)}$ but no pole at all for the dressed result $D_2$ since the anomalous dimension vanishes quadratically in this case. Therefore we do not pick that contribution from the $U(1)$ result. The contribution is given by $\mathcal{E}_2=-2\sigma\, \mathcal{C}_{2}(0)\,{\cal F}_{1,0}(\tau) $,
\beq
\mathcal{E}_2= 4\log \(\frac{u_1}{u_3}\)  \log(1-u_2)\ .\nn
\eeq
It is independent of $\phi$ and should be subtracted from $r_{U(1)}$ in $X_3$ (\ref{X3structure}). In other words we should remove the $\phi$ independent $\sigma$ linear contribution in $r_{U(1)}$.
\end{enumerate}
All together, we get that $(-2)X_3=-4r_{U(1)} +\mathcal{E}_1+\mathcal{E}_2$, i.e.
\beqa
X_3&=&
-4\[ \text{Li}_2\left(-\frac{1}{\mu  b_1}\right)+ \text{Li}_2\left(-\frac{\mu
   }{b_1}\right)\]-2\sum_{i=1}^3\text{Li}_2\left(1-u_i\right)\nn\\&&-2\log
   ^2\left(u_1\right)-2\log \left(1-u_2\right) \log \left(\frac{u_2}{u_1}\right)-2\log \left(1-u_3\right) \log
   \left(\frac{\left(1-u_2\right) u_3}{u_1}\right)+{2\pi^2\over3}\ .\nn
\eeqa

\subsection{More details on $Y_2$}\la{Y2appendix}

As argued in section \ref{Y2sec}, $Y_2$ takes the form
\beq
\text{Sym}[Y_2]= 2f_0 \otimes \mu +2\sum_{i=1}^3 f_i \otimes \(1+1/u_j\) \,.
\eeq
We will now proceed and find the functions $f_0$, $f_1$, $f_2$, $f_3$ by imposing is that $Y_2$ is the symbol of a function. To check that $Y_2$ is a symbol of a function we need to pick a set of independent variables. A very useful set of such variables is
\beq
x_j\in\{\mu,b_1,b_3\} \,.
\eeq
The $u_1 \leftrightarrow u_3$ symmetry of $Y_2$ leads to $f_1(\mu,b_1,b_3)=f_3(\mu,b_3,b_1)$.
The checks that $Y_2=a\otimes b \otimes c +a'\otimes b' \otimes c' \dots $ is the symbol of a function function read (\ref{symbolfunctioncheck})
\beqa
\!\!0\!\!&=&\!\!\( \partial_{j} \log a\, \partial_{k} \log b -\partial_{k} \log a \,\partial_{j} \log b \) \log c + \( \partial_{j} \log a' \,\partial_{k} \log b' - \partial_{k} \log a'\, \partial_{j} \log b' \) \log c'  + \dots   \la{eq1}\nn \\
\!\!0\!\!&=&\!\!\log a \( \partial_{j} \log b\, \partial_{k} \log c -\partial_{k} \log b\, \partial_{j} \log c \)  +\log a' \( \partial_{j} \log b' \,\partial_{k} \log c' - \partial_{k} \log b' \,\partial_{j} \log c' \)   + \dots  \nn
\eeqa
Each of these equations gives us several constraints since the coefficient of the different  logarithms in the right hand side of these equations  must be zero seperately. More explicitly we find
\beq
0=(\dots) \log b_1 + (\dots) \log(1+ b_1 \mu) + \dots
\eeq
for each equation. In total we find a set of  partial differential equations for the functions $f_a$. For example, the coefficient of $\log(1+b_1 \mu)$ of (\ref{eq1}) with $j,k=1,2$ yields
\beq
0=\left(\mu ^2-1\right) f_1\left(b_1,b_3,\mu \right)-\mu  \left(b_1 \mu +b_3 \mu +\mu ^2+1\right)
   \frac{\partial}{\partial {\mu} }f_1\left(b_1,b_3,\mu \right)
\eeq
which implies that
\beq
f_1(b_1,b_3,\mu)=\(b_1+b_3+\mu+\frac{1}{\mu}\) f_1(b_1,b_3)
\eeq
We pick this solution, plug it into the remaining differential equations, \verb"Simplify" them, pick the simplest differential equation, \verb"DeSolve" it, plug the solution again, etc.  At the end of the day, the eight differential equations that we found almost fix $Y_2$ completely!  We find
\beq
\text{Sym}[Y_2]=-2\[u_1\otimes \frac{1-u_1}{u_1}+u_3\otimes \frac{1-u_3}{u_3}+4 \mu \otimes \mu-f(u_2)\otimes \frac{1-u_2}{u_2}\]
\eeq
where $Y_2$ is a symbol of a function for any function $f_2(u_2)$. This function is fixed in the main text.

\section{Limiting values of $r(X,Y)$}
\label{limitingvalues}

The pairing $r_{U(1)}(X,Y)$ does not diverge if a single vertex of $X$ and a single vertex of $Y$ become null-separated, i.e.
 $X_a \cdot Y_c=0$ while all other space-time and twistor products are finite and non-zero.
 Indeed, the discontinuity $\Delta_{ac}$ is zero as  $X_a \cdot Y_c=0$. We can easily take the
limit  $X_a \cdot Y_c \to 0$ of the answer. It takes a little bit of work
if we use the expression in terms of space-time cross-ratios.
Indeed the argument of dilogarithms where  $X_a \cdot Y_c$ is in the numerator of the cross-ratio
go to $1$ in the limit, and we can drop them, but
the argument of  dilogarithms where  $X_a \cdot Y_c$ is in the denominator of the cross-ratio  goes to infinity, and we need to manipulate them further.
Up to constants, we can rewrite
\begin{equation}
-Li_2\left[1- \frac{(X_i \cdot Y_k)( X_{i-1}\cdot Y_{k+1})}{(X_i \cdot Y_{k+1})( X_{i-1}\cdot Y_{k})}\right]
\end{equation}
as
\begin{equation}
Li_2\left[1- \frac{(X_i \cdot Y_{k+1})( X_{i-1}\cdot Y_{k})}{(X_i \cdot Y_k)( X_{i-1}\cdot Y_{k+1})}\right]  + \frac{1}{2} \log^2 \frac{(X_i \cdot Y_k)( X_{i-1}\cdot Y_{k+1})}{(X_i \cdot Y_{k+1})( X_{i-1}\cdot Y_{k})}
\end{equation}
in order to bring $X_a \cdot Y_c$ in the numerator of the cross-ratio in the dilogarithm, which can then be dropped.
The squared logarithm terms collect to
\begin{equation}
  \frac{1}{2} \log^2 \frac{(X_a \cdot Y_{c-1})( X_{a-1}\cdot Y_{c})}{(X_a \cdot Y_{c})( X_{a-1}\cdot Y_{c-1})} + \frac{1}{2} \log^2 \frac{(X_{a+1} \cdot Y_c)( X_{a}\cdot Y_{c+1})}{(X_{a+1} \cdot Y_{c+1})( X_{a}\cdot Y_{c})}
\end{equation}
and must be combined with the terms in $r_2$ which contain $\log X_a \cdot Y_c$
 \begin{equation}
-  \log X_a \cdot Y_c \log \frac{(X_a \cdot Y_c)( X_{a+1}\cdot Y_{c+1})}{(X_a \cdot Y_{c+1})( X_{a+1}\cdot Y_{c}) }
-  \log X_{a-1} \cdot Y_{c-1} \log \frac{(X_{a-1} \cdot Y_{c-1})( X_{a}\cdot Y_{c})}{(X_{a-1} \cdot Y_{c})( X_{a}\cdot Y_{c-1}) }
\end{equation}
and
 \begin{equation}
-  \log X_a \cdot Y_{c-1} \log \frac{(X_a \cdot Y_{c-1})( X_{a+1}\cdot Y_{c})}{(X_{a} \cdot Y_{c})( X_{a+1}\cdot Y_{c-1}) }
-  \log X_{a-1} \cdot Y_c \log \frac{(X_{a-1} \cdot Y_c)( X_{a}\cdot Y_{c+1})}{(X_{a-1} \cdot Y_{c+1})( X_{a}\cdot Y_{c}) }
\end{equation}
The $\log^2 X_a \cdot Y_c$ terms cancel. So do the terms proportional to $\log X_a \cdot Y_c$.
What is left is a finite combination of logarithms.

On the other hand, if we use the $r^\pm$ expressions, the work is done for us: indeed none of the twistor inner products goes to zero,
all that happens in the limit is that the argument of some dilogarithms goes to $1$.

We will encounter a more intricate limit, where a vertex of one polygon is brought to the null line defined by two vertices of the
other polygon, see figure \ref{Wilsoncorrelator}(b).
Say that $X_a$ is brought to the line which passes through $Y_c$ and $Y_c+1$.
This is a more subtle limit, as many quantities go to zero at the same time:
$X_a \cdot Y_c$ and $X_a \cdot Y_{c+1}$, but also $(\lambda_a, \hat \mu_{c+1})$ and  $(\lambda_{a+1}, \hat \mu_{c+1})$,
and the dual products.

Let's consider the limit of $r^+(X,Y)$.
It is useful to express the limit as the combination of three scalar constraints.
Two are the obvious $(\lambda_a, \hat \mu_{c+1}) \to 0$ and  $(\lambda_{a+1}, \hat \mu_{c+1}) \to 0$.
They are equivalent to the $X_a \cdot \hat \mu_{c+1}=0$ constraint. A third constraint follows from
$X_a \wedge \mu_{c+1} =0$ (notice the lack of hat!), and takes the form $X_a \cdot \left( \hat \mu_c \wedge \hat \mu_{c+2}\right)\to0$,
or $ (\lambda_a, \hat \mu_{c})  (\lambda_{a+1}, \hat \mu_{c+2}) \to  (\lambda_{a+1}, \hat \mu_{c})  (\lambda_{a}, \hat \mu_{c+2})$.
Generically, none of these four products go to zero, and hence their cross-ratio will go to $1$.
\begin{equation}
\frac{(\lambda_a, \hat \mu_{c})  (\lambda_{a+1}, \hat \mu_{c+2}) }{  (\lambda_{a+1}, \hat \mu_{c})  (\lambda_{a}, \hat \mu_{c+2})}\to 1
\end{equation}

It is interesting to look at the simplification of the following crucial ratio
\begin{equation}
t= \frac{X_a \cdot Y_{c+1} }{X_a \cdot Y_c }=\frac{(\lambda_a, \hat \mu_{c+1})(\lambda_{a+1}, \hat \mu_{c+2})-(\lambda_{a+1}, \hat \mu_{c+1})(\lambda_a, \hat \mu_{c+2})}{(\lambda_a, \hat \mu_{c})(\lambda_{a+1}, \hat \mu_{c+1})-(\lambda_a, \hat \mu_{c+1})(\lambda_{a+1}, \hat \mu_{c})}\end{equation}
Due to the third constraint, we have
\begin{equation}
t \to -\frac{(\lambda_{a+1}, \hat \mu_{c+2})}{(\lambda_{a+1}, \hat \mu_{c})} \qquad t\to-\frac{(\lambda_{a}, \hat \mu_{c+2})}{(\lambda_{a}, \hat \mu_{c})}
\end{equation}
There is a simple geometric meaning to the ratio $t$ which enters the limit. The point $X_a$ in the limit can be written as $Y_{c+1} - t Y_c$.

If we look at the explicit expression for $r^+$,
we have some dilogarithms whose argument goes to zero
\begin{equation}
 - Li_2\left[ \frac{(\lambda_{a+1}, \hat \mu_{c+1})(\lambda_{a+2}, \hat \mu_{c})}{(\lambda_{a+1}, \hat \mu_{c})(\lambda_{a+2}, \hat \mu_{c+1})}\right]
-Li_2\left[ \frac{(\lambda_{a-1}, \hat \mu_{c+2})(\lambda_{a}, \hat \mu_{c+1})}{(\lambda_{a-1}, \hat \mu_{c+1})(\lambda_{a}, \hat \mu_{c+2})}\right]\end{equation}
Some other depend on the ratio $(\lambda_{a+1}, \hat \mu_{c+1})/(\lambda_{a}, \hat \mu_{c+1})$
\begin{equation}
 -  Li_2\left[ \frac{(\lambda_a, \hat \mu_{c+1})(\lambda_{a+1}, \hat \mu_{c})}{(\lambda_a, \hat \mu_{c})(\lambda_{a+1}, \hat \mu_{c+1})}\right]
 -Li_2\left[ \frac{(\lambda_a, \hat \mu_{c+2})(\lambda_{a+1}, \hat \mu_{c+1})}{(\lambda_a, \hat \mu_{c+1})(\lambda_{a+1}, \hat \mu_{c+2})}\right]
\end{equation}
but due to the third constraint, their product of their arguments go to $1$, and we can replace them with
\begin{equation}
\frac{1}{2} \log^2 \left[ \frac{(\lambda_a, \hat \mu_{c+1})(\lambda_{a+1}, \hat \mu_{c})}{(\lambda_a, \hat \mu_{c})(\lambda_{a+1}, \hat \mu_{c+1})}\right]
\end{equation}
Finally, we have some dilogarithms whose argument diverge,
\begin{equation}
-Li_2\left[ \frac{(\lambda_{a-1}, \hat \mu_{c+1})(\lambda_{a}, \hat \mu_{c})}{(\lambda_{a-1}, \hat \mu_{c})(\lambda_{a}, \hat \mu_{c+1})}\right]
- Li_2\left[ \frac{(\lambda_{a+1}, \hat \mu_{c+2})(\lambda_{a+2}, \hat \mu_{c+1})}{(\lambda_{a+1}, \hat \mu_{c+1})(\lambda_{a+2}, \hat \mu_{c+2})}\right]
\end{equation}
and need to be manipulated by the replacement $Li_2(1/x) \to -Li_2(x) - \frac{1}{2} \log^2 x$,
giving
\begin{equation}
\frac{1}{2}\log^2 \left[ \frac{(\lambda_{a-1}, \hat \mu_{c+1})(\lambda_{a}, \hat \mu_{c})}{(\lambda_{a-1}, \hat \mu_{c})(\lambda_{a}, \hat \mu_{c+1})}\right]
+\frac{1}{2} \log^2\left[ \frac{(\lambda_{a+1}, \hat \mu_{c+2})(\lambda_{a+2}, \hat \mu_{c+1})}{(\lambda_{a+1}, \hat \mu_{c+1})(\lambda_{a+2}, \hat \mu_{c+2})}\right]
\end{equation}

Now we need to combine all those squared logarithms with $r^+_2$. Thanks again to the third constraint, all terms containing either
$\log (\lambda_a, \hat \mu_{c+1})$ or $\log (\lambda_{a+1}, \hat \mu_{c+1})$ cancel out, and the result has a finite limit.

Hence the correct limit for $r^+$ is obtained by adding together the main pieces
\begin{equation}
r^+_{1'} = - \sum'_{i,k} Li_2\left[ \frac{(\lambda_i, \hat \mu_{k+1})(\lambda_{i+1}, \hat \mu_{k})}{(\lambda_i, \hat \mu_{k})(\lambda_{i+1}, \hat \mu_{k+1})}\right]
\end{equation}
\begin{equation}
r^+_{2'} =  \sum'_{i,k} \log (\lambda_i, \hat \mu_{k}) \log \frac{(\lambda_i, \hat \mu_{k+1})(\lambda_{i+1}, \hat \mu_{k})}{(\lambda_i, \hat \mu_{k})(\lambda_{i+1}, \hat \mu_{k+1})}
\end{equation}
where each of the sum omits the six terms where $i=a-1$ or $i=a$ or $i=a+1$ and $k=c$ or $k=c+1$,
and the remaining logs
\begin{equation}
r^+_3 = \log (\lambda_{a-1}, \hat \mu_{c+1})\log \frac{(\lambda_{a-1}, \hat \mu_{c+2})(\lambda_{a}, \hat \mu_{c})}{(\lambda_a, \hat \mu_{c+2})(\lambda_{a-1}, \hat \mu_{c})}+\log (\lambda_{a+2}, \hat \mu_{c+1})\log \frac{(\lambda_{a+1}, \hat \mu_{c+2})(\lambda_{a+2}, \hat \mu_{c})}{(\lambda_{a+2}, \hat \mu_{c+2})(\lambda_{a+1}, \hat \mu_{c})}
\end{equation}
and
\begin{equation}
r^+_4 =
-\frac{1}{2} \log^2 \frac{(\lambda_{a-1}, \hat \mu_{c})}{(\lambda_{a-1}, \hat \mu_{c+1})}-\frac{1}{2} \log^2\frac{(\lambda_{a+1}, \hat \mu_{c})}{(\lambda_{a+2}, \hat \mu_{c})}+\frac{1}{2} \log^2  \frac{(\lambda_{a+1}, \hat \mu_{c+2})}{(\lambda_{a+2}, \hat \mu_{c+2})}+\frac{1}{2} \log^2  \frac{(\lambda_{a+2}, \hat \mu_{c})}{(\lambda_{a+2}, \hat \mu_{c+1})}
\end{equation}

The condition \begin{equation}
\frac{(\lambda_a, \hat \mu_{c})  (\lambda_{a+1}, \hat \mu_{c+2}) }{  (\lambda_{a+1}, \hat \mu_{c})  (\lambda_{a}, \hat \mu_{c+2})}\to 1
\end{equation}
must be true for rescaling invariance to hold.

\end{document}